# Investigating Protostellar Accretion-driven Outflows across the Mass Spectrum: JWST NIRSpec Integral Field Unit 3–5 μm Spectral Mapping of Five Young Protostars

Samuel A. Federman[1] , S. Thomas Megeath[1] , Adam E. Rubinstein[2] , Robert Gutermuth[3] , Mayank Narang[4,5] ,
Himanshu Tyagi[5] , P. Manoj[5] , Guillem Anglada[6] , Prabhani Atnagulov[1] , Henrik Beuther[7] , Tyler L. Bourke[8] ,
Nashanty Brunken[9] , Alessio Caratti o Garatti[10] , Neal J. Evans, II[11] , William J. Fischer[12] , Elise Furlan[13] ,
Joel D. Green[12] , Nolan Habel[14] , Lee Hartmann[15] , Nicole Karnath[16,17] , Pamela Klaassen[18] , Hendrik Linz[7,19] ,
Leslie W. Looney[20,21] , Mayra Osorio[6] , James Muzerolle Page[12] , Pooneh Nazari[9] , Riwaj Pokhrel[1] ,
Rohan Rahatgaonkar[22] , Will R. M. Rocha[23] , Patrick Sheehan[24] , Katerina Slavicinska[9] , Thomas Stanke[25] ,
Amelia M. Stutz[26] , John J. Tobin[21] , Lukasz Tychoniec[9] , Ewine F. Van Dishoeck[9,25] , Dan M. Watson[2] ,
Scott Wolk[17] , and Yao-Lun Yang[27]

[1] Ritter Astrophysical Research Center, Dept. of Physics and Astronomy, University of Toledo, Toledo, OH 43606, USA
[2] University of Rochester, Rochester, NY 14627, USA
[3] University of Massachusetts Amherst, MA 01003, USA
[4] Institute of Astronomy and Astrophysics Academia Sinica, Taiepei, Taiwan
[5] Tata Institute of Fundamental Research, Mumbai, Maharashtra, India
[6] Instituto de Astrofísica de Andalucía, CSIC, Glorieta de la Astronomía s/n, E-18008 Granada, Spain
[7] Max Planck Institute for Astronomy, Heidelberg, Baden Wuerttemberg, Germany
[8] SKA Observatory, Jodrell Bank, Lower Withington, Macclesfield SK11 9FT, UK
[9] Leiden Observatory, Universiteit Leiden, Leiden, Zuid-Holland, The Netherlands
[10] INAF-Osservatorio Astronomico di Capodimonte, Italy
[11] Department of Astronomy, The University of Texas at Austin, 2515 Speedway, Stop C1400, Austin, TX 78712-1205, USA
[12] Space Telescope Science Institute, 3700 San Martin Drive, Baltimore, MD 21218, USA
[13] Caltech/IPAC, Pasadena, CA 91125, USA
[14] Jet Propulsion Laboratory, Pasadena, CA 91011, USA
[15] University of Michigan, Ann Arbor, MI 48109, USA
[16] Space Science Institute, Boulder, CO 80301, USA
[17] Center for Astrophysics | Harvard & Smithsonian, Cambridge, MA 02138, USA
[18] United Kingdom Astronomy Technology Centre, Edinburgh, UK
[19] AG Analytische Mineralogie, Friedrich-Schiller-Universität, Jena,Thüringen, Germany
[20] Department of Astronomy, University of Illinois, 1002 West Green Street, Urbana, IL 61801, USA
[21] National Radio Astronomy Observatory, Charlottesville, VA 22903, USA
[22] Gemini South Observatory, La Serena, Chile
[23] Laboratory for Astrophysics, Leiden Observatory, Universiteit Leiden, Leiden, Zuid-Holland, The Netherlands
[24] Northwestern University, Evanston, IL 60208, USA
[25] Max-Planck Institut für Extraterrestrische Physik, Garching bei München, Germany
[26] Departamento de Astronomía, Universidad de Concepción, Casilla 160-C, Concepción, Chile
[27] Star and Planet Formation Laboratory, RIKEN Cluster for Pioneering Research, Wako, Saitama 351-0198, Japan

## Abstract

Investigating Protostellar Accretion is a Cycle 1 JWST program using the NIRSpec+MIRI integral field units to obtain 2.9–28 μm spectral cubes of five young protostars with luminosities of 0.2–10,000 $L_\odot$ in their primary accretion phase. This paper introduces the NIRSpec 2.9–5.3 μm data of the inner 840–9000 au with spatial resolutions from 28 to 300 au. The spectra show rising continuum emission; deep ice absorption; emission from $H_2$, H I, and [Fe II]; and the CO fundamental series in emission and absorption. Maps of the continuum emission show scattered light cavities for all five protostars. In the cavities, collimated jets are detected in [Fe II] for the four <320 $L_\odot$ protostars, two of which are additionally traced in Brα. Knots of [Fe II] emission are detected toward the most luminous protostar, and knots of [Fe II] emission with dynamical times of <30 yr are found in the jets of the others. While only one jet is traced in $H_2$, knots of $H_2$ and CO are detected in the jets of four protostars. $H_2$ is seen extending through the cavities, showing that they are filled by warm molecular gas. Bright $H_2$ emission is seen along the walls of a single cavity, while in three cavities narrow shells of $H_2$ emission are found, one of which has an [Fe II] knot at its apex. These data show cavities containing collimated jets traced in atomic/ionic gas surrounded by warm molecular gas in a wide-angle wind and/or gas accelerated by bow shocks in the jets.

## 1. Introduction

Stars accumulate most of their mass during the protostellar phase through the infall and accretion of interstellar gas and dust. This phase is characterized by a rapidly evolving, infalling envelope surrounding the nascent protostar that is consumed and dispersed on a ~500,000 yr timescale (Dunham et al. 2014). This process is mediated by infall onto a circumstellar disk from which mass is accreted onto the protostar. The accretion drives powerful outflows in the form of collimated jets and wide-angle winds, clearing out cavities from the dense envelope. Up to 80% of the total stellar mass is accumulated during the primary accretion phase when the protostar is still embedded in a dense envelope and has a high accretion rate (Fischer et al. 2017; Federman et al. 2023).

Because protostars in their primary accretion phase are deeply embedded within dense envelopes, they are obscured at visible and shorter wavelengths. Observing protostellar growth, therefore, requires near-infrared (NIR) wavelengths and longer. After the end of the Spitzer mission, observations in the IR were mostly restricted to ground-based facilities, limiting the sensitivity of IR observations at $>3\,\mu$m. JWST, which launched on 2021 December 25 and reached its final orbit at L2 on 2022 January 24, has sparked a revolution in NIR to mid-IR astronomy (Böker et al. 2023; Rigby et al. 2023; Wright et al. 2023). The wavelength coverage, angular and spectral resolution, and unparalleled sensitivity of JWST are perfectly suited for studying star formation in the innermost regions of protostars. JWST's ability to detect warm gas in accretion flows and outflow shocks in these regions complements the ability of the Atacama Large Millimeter/submillimeter Array (ALMA) to map the cold gas component with a similar angular resolution.

Investigating Protostellar Accretion (IPA) is a Cycle 1 medium GO program on JWST (PID 1802; Megeath et al. 2021). Using the NIRSpec+MIRI integral field units (IFUs) to obtain 2.9–28 $\mu$m spectral cubes, IPA targets protostars with moderate to high disk inclinations in their primary accretion phase with luminosities of 0.2–10,000 $L_{\odot}$ and central masses of 0.12–12 $M_{\odot}$. In this primary phase, there is little direct information on the accretion process. This phase of protostellar growth is characterized by strong, accretion-powered outflows in the form of collimated jets and wide-angle winds. It is through the interaction of jets and winds launched by the central protostar that a cavity is cleared in the envelope. Along the surfaces of the cavity, infalling gas is entrained in the outflow through the transfer of momentum.

Despite the importance of infall and outflow for determining the masses of the stars and the star formation efficiency, the processes by which outflows clear cavities and entrain gas are still debated. Cavities can be carved by high-velocity X-winds, high-velocity pulsed jets, or lower-velocity disk winds (e.g., Machida & Hosokawa 2013; Shang et al. 2020; Ray & Ferreira 2021; Rabenanahary et al. 2022; Rivera-Ortiz et al. 2023). The IPA program will search for signatures of accretion in emission and absorption lines, map the flows of gas inside outflow cavities using shock tracers, and examine the interaction between outflows and the surrounding cavity (Narang et al. 2024; Rubinstein et al. 2023a; D. Watson et al. 2024, in preparation). The program will also characterize the chemical and spatial variation of ices in the surrounding envelope through their absorption features (Brunken et al. 2024; Nazari et al. 2024; K. Slavicinska et al. 2024, in preparation; H. Tyagi et al. 2024, in preparation). With a roughly logarithmic distribution in protostellar mass and luminosity (Table 1), we will evaluate how accretion, outflow, and ices vary with these properties.

In this paper, we present the first overview of the five IPA sources in the NIRSpec range (2.9–5.3 $\mu$m). These data show rich line emission from outflows, in particular emission lines due to rotational transitions of $H_2$, the CO rovibrational fundamental series, H I, and [Fe II]. With spatial resolutions down to 30 au (depending on the distance to the protostar), the observed emission lines trace the shocks in jets and along the cavity walls, providing a detailed portrait of how these young stars interact with their infalling envelopes. We present representative NIRSpec spectra for the protostars. We then display the morphologies for each protostar in the continuum and key emission lines at 2.9–5.3 $\mu$m. We explore how the spectra spatially vary and discuss the physical origins of the emission lines and their implications for protostellar outflows.

## 2. Sample and Observations

The IPA program targets five protostars whose basic properties have been characterized by existing observations, including established heliocentric distances, well-sampled mid-IR to far-IR spectral energy distributions (SEDs), and interferometric observations of their disks. From these data, the targets were selected by criteria including proximity to the Sun for a given luminosity (140 pc–1.6 kpc); bolometric luminosities between 0.2 and $10^4\,L_{\odot}$ in a roughly geometric progression; protostellar masses from 0.12 to 12 $M_{\odot}$ in a roughly geometric progression; and moderate to high, but not edge-on, inclinations. The bias of our sample to higher (65°–84°) inclinations is advantageous for observing protostellar outflows, but at these inclinations the bolometric luminosities typically underestimate the total luminosities of the protostars (Fischer et al. 2017). Although the protostars were also selected on the lack of evidence for significant luminosity variability at the time of the sample definition, the variability of protostars limits our knowledge of the luminosity at the time of the JWST observations (e.g., B335-IRS; Evans et al. 2023). The central protostellar masses have been determined from Keplerian motions in the disk or from model fitting to the envelope properties (in the case of B335-IRS; Evans et al. 2023). We also chose protostars that appear to be in their primary mass accretion phase. The three low-mass objects IRAS 16253–2429 (hereafter IRAS 16253), B335-IRS (hereafter B335), and HOPS 153 are Class 0 protostars. The intermediate-mass/luminosity protostar HOPS 370 has a $T_{bol} = 71.5$ K, close to the Class 0/I boundary of $T_{bol} = 70$ K, and a high envelope 870 $\mu$m flux, indicating that it is still deeply embedded (Furlan et al. 2016; Federman et al. 2023). The high-mass protostar IRAS 20126+4104 (hereafter IRAS 20126) exhibits an SED similar to low-mass Class 0 protostars. Finally, we selected protostars that appear to be forming single stars primarily based on the lack of companions in ALMA data. The one exception is the most massive protostar IRAS 20126, which has two likely companions (Section 3); this source was targeted since almost all massive stars have companions. Adopted properties for each protostar are listed in Table 1, along with their references.

The NIRSpec observations were taken between 2022 July 22 and October 16. The targets were all observed in the G395M ($R \sim 1000$) setting in a $2 \times 2$ mosaic with 10% overlap and a four-point dither pattern. The NIRSpec IFU has a field of view (FOV) of $3'' \times 3''$, and the resulting mosaics cover a total FOV of $6'' \times 6''$. We used the NRSIRS2RAPID readout mode with a single exposure of 30 groups and one integration (7003 s exposure time), two exposures of 15 groups and one

**Table 1**
Source Properties for the IPA Sample

| Parameter | IRAS 16253 | B335 | HOPS 153 | HOPS 370 | IRAS 20126 | References |
|---|---|---|---|---|---|---|
| R.A. | 16:28:21.615 | 19:37:00.89 | 5:37:57.021 | 5:35:27.635 | 20:14:26.0364 | (1, 2, 3, 4, 5) |
| Decl. | −24:36:24.33 | +7:34:09.6 | −7:06:56.25 | −5:09:34.44 | +41:13:32.516 | (1, 2, 3, 4, 5) |
| Distance (pc) | 140 | 165 | 390 | 390 | 1550 | (6, 7, 3, 3, 8) |
| $L_{bol}$ ($L_\odot$) | 0.16 | 1.4[a] | 3.8 | 315.7 | $10^4$ | (1, 9, 3, 3, 10) |
| $T_{bol}$ (K) | 42 | 37 | 39.4 | 71.5 | 60[b] | (11, 12, 13, 13, 14) |
| Stellar mass ($M_\odot$) | 0.12–0.17 | 0.25 | 0.6 | 2.5 | 12 | (1, 9, 15, 4, 10) |
| Disk radius (au) | 13–19 | <10 | 150 | 100 | 860 | (1, 18, 3, 3, 10) |
| Disk inclination (deg) | 65 | 83.5 | 74.5 | 72.2 | 82 | (19, 9/16, 3, 3, 17) |

**Notes.**
[a] Observed luminosity of B335 underwent a period of luminosity variation on the order of factors of 5–7 from 2010 to 2015, but it has since returned to its previous luminosity (Evans et al. 2023).
[b] The temperature for IRAS 20126 is not $T_{bol}$ but rather the temperature of a graybody fit to the continuum SED in Cesaroni et al. (1999).
**References.** (1) Aso et al. 2023; (2) Maury et al. 2018; (3) Tobin et al. 2020a; (4) Tobin et al. 2020b; (5) Cesaroni et al. 2014; (6) Ortiz-León et al. 2018; (7) Watson 2020; (8) Reid et al. 2019; (9) Evans et al. 2023; (10) Chen et al. 2016; (11) Hsieh et al. 2015; (12) Andre et al. 2000; (13) Furlan et al. 2016; (14) Cesaroni et al. 1999; (15) J. J. Tobin 2024, private communication; (16) Bjerkeli et al. 2019; (17) Massi et al. 2023; (18) Yen et al. 2015; (19) Yen et al. 2017.

integration (3501 s), or a single exposure of six groups and five integrations (7003 s), depending on the source. The observing techniques were selected to maximize signal without saturation for each source. Leak calibration measurements were performed on the three highest-mass sources (HOPS 153, HOPS 370, and IRAS 20126), which have more complex surrounding environments. A summary of the observations for each source is in Table 2.

### 2.1. Data Reduction Method

In lieu of the automated pipeline reduction products provided through the Mikulski Archive for Space Telescopes (MAST), we performed local reductions to the raw data using the publicly available JWST pipeline software (Bushouse et al. 2023) to customize the NIRSpec data treatment in several ways. This was necessary because, especially in the early stages of JWST, the standard pipeline products had a number of artifacts, warm pixels, and other issues that had to be addressed through custom reduction. We used the CRDS version 11.16.20 to generate the cubes for this work. Stage 1, which applies corrections to the raw data and computes the "rate" images from the raw samples up the ramp using nondestructive reads, is run as normal, with no customizations other than the inclusion of up-to-date "pmap" calibration files and settings. After this stage is applied, we flag pixels with data quality values 2097152 and 2097156 as "DO_NOT_USE."

For Stage 2, which applies various physical corrections to properly calibrate the individual "rate" images in both flux and wavelength to make "cal" image products, we disabled the "wavecorr" and "pathloss" corrections but otherwise ran the rest of the stage as normal. We inspected the data and saw no evidence of artifacts from light leaks; thus, we disabled the application of the leak calibration imprint subtraction for the sources where we took those observations in order to avoid extra noise from subtracting those spectra.

After Stage 2 is complete, we carefully crop out apparently bad pixels, many of which appear to be transient from one target to the next yet are reasonably consistent across recent dithers for a given target. Most bad pixels are not flagged and removed by the outlier detection and removal step in Stage 3. That step was destructive to the valid parts of our data; thus, we disable it during Stage 3. Instead, we use the Stage 2 data products to flag out transient bad pixels with relatively little loss of desirable data. We describe this custom masking process in detail in Appendix A. In addition to those custom masked pixels, we also set the following data quality flags as "DO_NOT_USE" at this point: 1024, 1026, 1028, 1030, 4195328, 4195330, 4195332, 134217728, 138412032, 138412036, 536871936, 536871938, and 536871940. In the early stages of the standard pipeline only a small subset of data quality flags were marked as "DO_NOT_USE," despite many others also being unrecoverable. In our custom processing of the data we set these flags to "DO_NOT_USE" based on the recommendation of experts at STScI. We also mask out extreme values in the science data themselves, removing any pixels with values $>10^5$ or $<-10^3$. Finally, we combine the images to make the final combined spectral cubes via Stage 3 using the default settings, except for disabling the outlier detection step as mentioned above. The astrometric calibration process for the NIRSpec and MIRI data is described in Appendix B.

**Table 2**
NIRSpec Observing Information for the IPA Targets

| | IRAS | | HOPS | | IRAS |
| Parameter | 16253 | B335 | 153 | HOPS 370 | 20126 |
|---|---|---|---|---|---|
| Obs date | 2022-07-22 | 2022-09-15 | 2022-09-24 | 2022-10-16 | 2022-08-09 |
| $N_{group}/N_{int}$ | 30/1 | 30/1 | 30/1 | 2 × 15/1 | 6/5 |
| Exp. Time (s) | 7003 | 7003 | 7003 | 2 × 3501 | 7003 |

### 3. Results

The IPA NIRSPEC IFU observations have produced five rich data cubes spanning 2.9–5.3 μm. In this section, we provide an initial examination of these cubes, each of which has footprints of $6'' × 6''$ with $0''.1$ pixels. The measured angular resolution of the NIRSpec data ranges from $0''.17$ to $0''.21$ across the wavelength coverage. We begin our examination with representative spectra of the central regions of the protostars. We then explore each protostar through line maps of bright spectral components: Brα, [Fe II], H₂ (0,0) S(11), CO $v = 1-0$ rovibrational transitions, and the scattered light continuum. Generally, Brα and other H I lines trace accretion on the smallest scales and shocks in outflows at larger distances from protostars (e.g., Komarova & Fischer 2020; Rubinstein et al. 2023b), while the [Fe II] and $H_2$ lines trace shocks in outflows (e.g., Stapelfeldt et al. 1991; Zinnecker et al. 1998; Reipurth et al. 2000). The CO emission is known to arise from both inner disks and outflows (e.g., Herczeg et al. 2011; Salyk et al. 2011). We then provide additional spectra from jet and scattered light positions to examine spatial variations of the spectra around each protostar.

### 3.1. Spectra of Central Regions

In Figure 1, we show the NIRSpec IFU spectra from the central regions of each protostar, extracted from a $0\farcs6$-radius circular aperture to obtain high signal-to-noise ratio (S/N) representative spectra of the central regions. The apertures are centered on the submillimeter/millimeter positions listed in Table 1. Despite the 5 mag range in luminosity, the spectra are strikingly similar, particularly for the protostars with luminosities $\leqslant 310 L_\odot$. Each spectrum shows several common elements: a rising continuum with deep ice absorption features, emission in ions, H I, $H_2$, and a forest of CO lines. The prominent ionic lines are primarily from [Fe II]. The most prominent and common H I line with clear detections outside of ice absorption features is the Brα line at 4.052 μm. A number of $H_2$ lines are detected; the (0,0) $S(11)$ at 4.181 μm is both prominent and uncontaminated by other lines and features. Finally, bright forests of the rovibrational transitions of the CO fundamental series are apparent; these include the $v = (1-0)$ $R$- and $P$-branch lines extending from 4.5 to 5.25 μm. The high-mass/luminosity protostar IRAS 20126 shows distinct differences, particularly at $\lambda > 5$ μm. In this wavelength range, the central position shows a much steeper rise in the spectrum. Furthermore, the emission from the CO fundamental series in the $R$-branch and part of the $P$-branch is weaker toward the central protostar, and the $P$-branch is detected in absorption beyond 5 μm. In Appendix C we list the lines detected in the central apertures, marked by the colored dashes at the top of the spectra in Figure 1.

### 3.2. Spatial Structure of the Lines and Continuum

The NIRSpec IFU spectra vary significantly with position across each protostellar field. To characterize these variations and the structures they delineate, we show maps of selected lines and continuum in Figures 2–6. In each map, we display strong lines that are least contaminated by other lines or ice features. The atomic and ionic lines are the Brα line (4.052 μm) and one of two [Fe II] lines: the 4.115 μm or the 4.889 μm line. Although the 4.889 μm [Fe II] line is the strongest for the lower-mass protostars, it is difficult to separate from the CO lines and bright continuum for the high-mass protostar IRAS 20126. We instead use the 4.115 μm [Fe II] line for IRAS 20126, which is outside of the CO forest. The molecular lines are the $H_2$ (0,0) $S(11)$ line at 4.181 μm and the sum of the CO $v = (1-0)$ $P(11)$–$P(18)$ lines (4.76–4.84 μm); we used the $P$-branch lines since they overlap with fewer other lines and the wings of the $R$-branch lines are more blended, making baseline subtraction difficult. We sum the CO lines to create a stacked image with a higher S/N than any individual line. Our line and image extraction subtracts a smoothed baseline at each spaxel and simultaneously fits line profiles for a library of species (see Rubinstein et al. (2023a) for details). For the

[Fe II] 4.115 μm map of IRAS 20126, we opt to subtract a linear fit to the local continuum to better account for blueshifted regions. The 5 μm continuum is determined from a linear fit to the minimum points of the data between 4.75 and 5.25 μm, masking out bright line emission. For the CO maps we made a linear fit to the minima between the CO lines to estimate the continuum. The spectral resolution of our data around 5 μm ($R \sim 1200$) is insufficient to fully separate the wings of the individual CO lines, potentially resulting in a slight overestimation of the continuum from the blending of the line wings, but this effect should be small.

#### 3.2.1. IRAS 16253

The lowest-luminosity and lowest-mass protostar in our sample is IRAS 16253, with a relatively small disk (13 au radius). The NIRSpec data resolve an 840 au diameter region centered on this protostar with 28 au spatial resolution, revealing the bipolar, hourglass morphology first seen at much larger scales in Spitzer IRAC imaging (Barsony et al. 2010). Figure 2 shows the narrow point of bright emission bridging the two cavities responsible for its designation as the wasp-waist nebula. This morphology continues down to the scales probed by JWST and is most apparent in the wide-angle 5 μm scattered light continuum (bottom left panel of Figure 2). The northern scattered light cavity is brighter than the southern cavity, indicating that the protostar is inclined with the northern cavity toward the observer. The higher spectral resolution MIRI data confirm that the northern jet is blueshifted relative to the southern jet (Narang et al. 2024). This inclination is also consistent with millimeter observations of molecular outflows coming from the source (e.g., Stanke et al. 2006; Hsieh et al. 2016; Aso et al. 2023). The fainter southern cavity shows three arcs, suggesting a more complex, 3D structure; two arcs trace wide angles, while the third appears to follow the eastern wall of the inner $H_2$ shell.

Inside the cavity, the Brα emission (top left panel of Figure 2) and [Fe II] (top right panel) are concentrated in a tightly collimated jet extending over 200 au in the northern cavity. The jet detected in [Fe II] in the southern cavity is more extinguished, in part due to the inclination of the northern cavity toward the observer (Narang et al. 2024). A bright knot is apparent in [Fe II] and Brα at the end of a gap in the northern jet. This knot is wider and slightly offset to the east from the rest of the jet. The Brα extends all the way down to the source, tracing both the jet and potentially accretion lines from the central protostar.

In the northern cavity, the warm $H_2$ component (middle left panel) traces a narrow shell-like structure that is located within the wider outflow cavity apparent in the continuum, as seen in the color-composite image (bottom right panel). At the tip of the northeastern limb, this $H_2$ structure appears to connect to the [Fe II]/Brα knot. The northern outflow cavity, which is inclined toward the observer, is fainter in warm $H_2$ than the more obscured southern outflow cavity. Investigating why the $H_2$ in the southern cavity inclined away from us is brighter than the northern $H_2$ is beyond the scope of this paper. Possible explanations include different excitation/temperature conditions in the outflow lobes, perhaps due to less oblique wide-angle wind shocks in the southern cavity. Future analysis will leverage the higher spectral resolution and additional $H_2$ lines available with MIRI to test the excitation conditions of the different outflow lobes. The southern cavity shows bright $H_2$ emission that follows the inner cavity walls apparent in the continuum, with fainter emission filling the cavity. The CO *P*-branch emission (middle right panel) is concentrated toward the base of the outflow. This emission fills the base of the northern cavity and is wider in extent than the jet emission in Brα. In the southern cavity, faint CO emission is detected along the inner cavity walls. In the wide-angle continuum image (bottom left panel) a faint background star is seen toward the northern cavity; a background-subtracted spectrum for this star is reported in Appendix D.

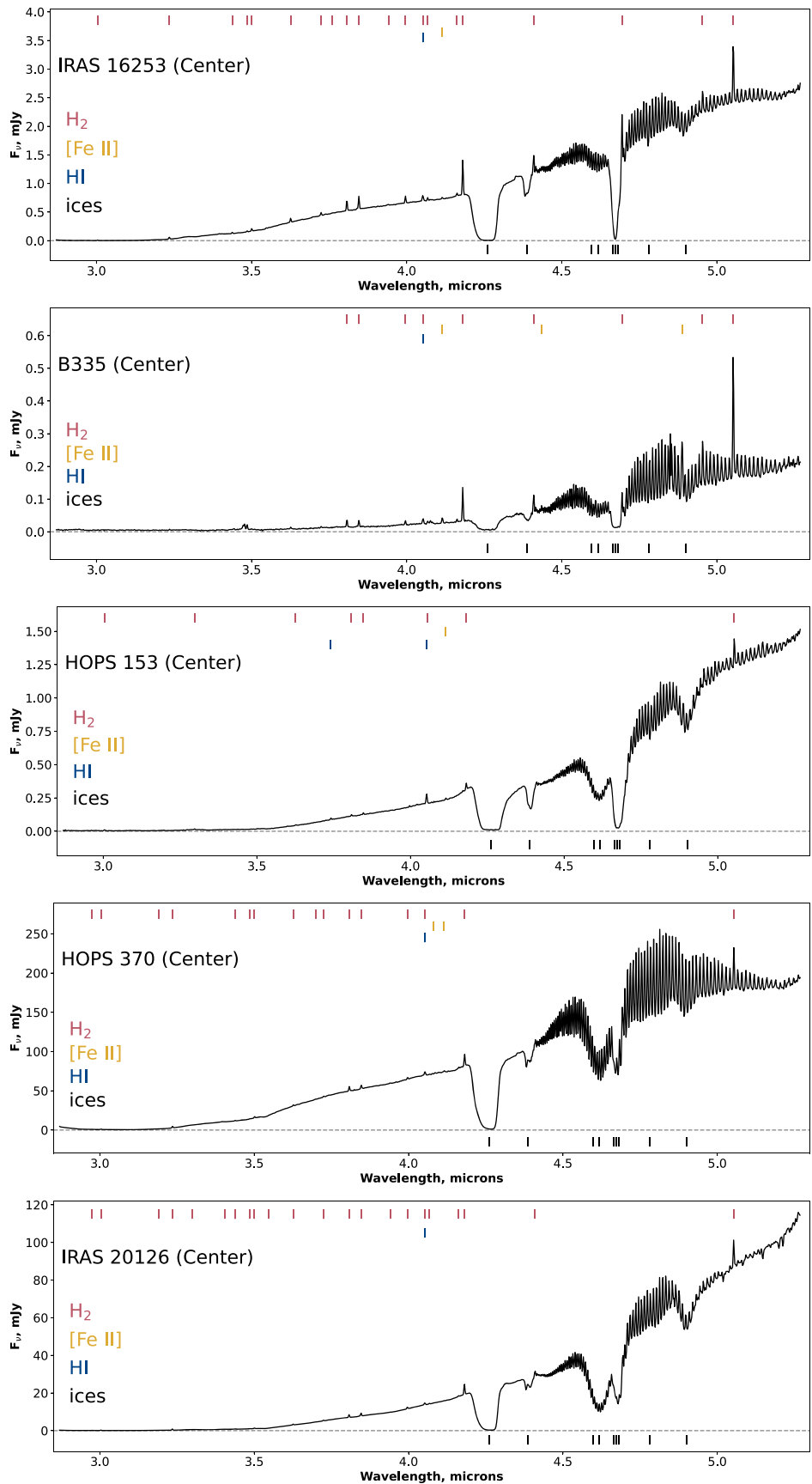

**Figure 1.** NIRSpec IFU central spectra for the five protostars. From top to bottom, increasing in protostellar mass: IRAS 16253, B335, HOPS 153, HOPS 370, and IRAS 20126. The colored dashes across the top denote the wavelengths of $H_2$, atomic, and ionic emission lines, and ice absorption features are denoted by black dashes on the bottom. The spectra are integrated over a 0″.6-radius circular aperture with no background subtraction.

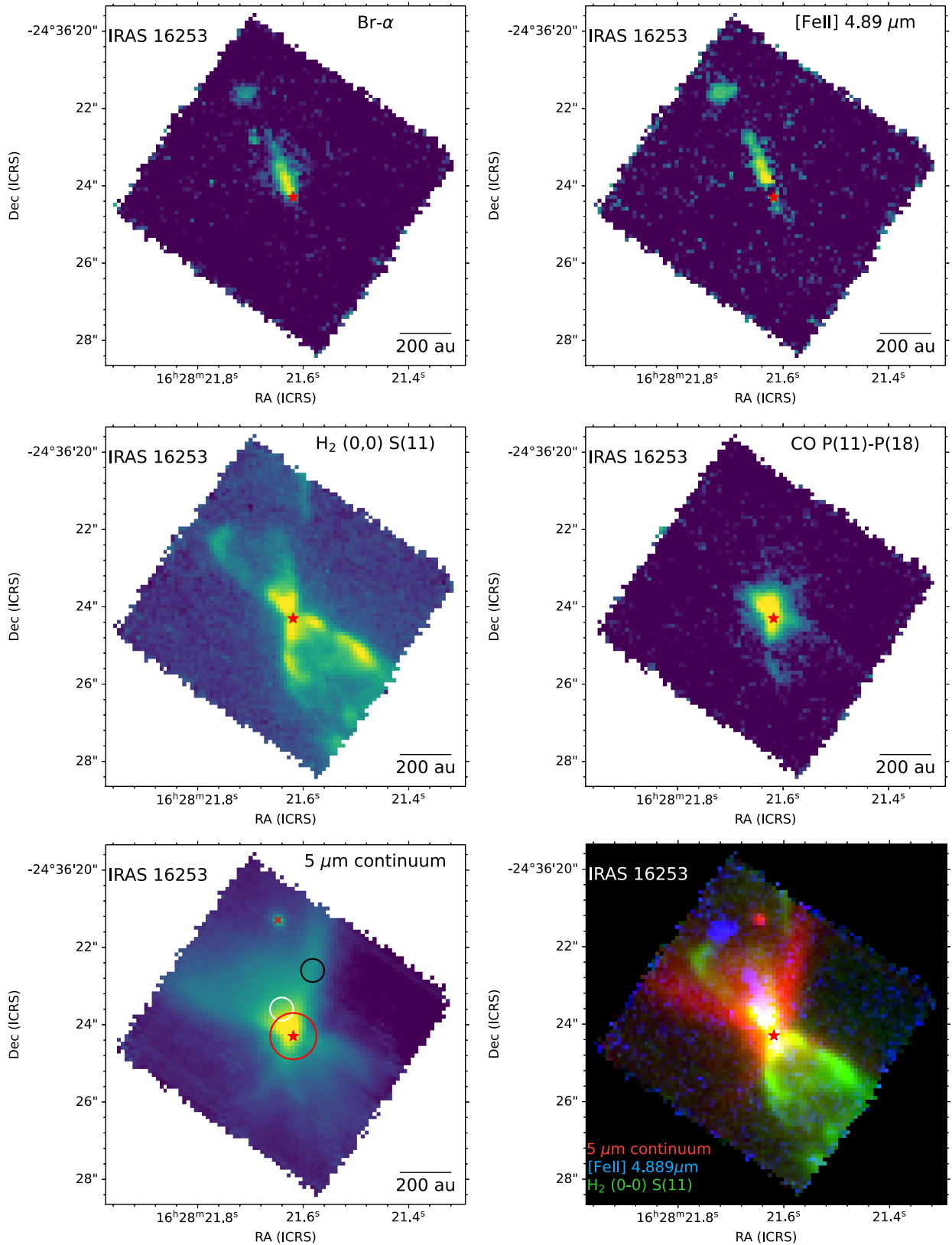

**Figure 2.** Selected line maps for IRAS 16253, displaying the different morphologies based on outflow tracer. All images are displayed with an arcsinh scale. Top: Brα (4.052 μm) and [Fe II] (4.889 μm) on the left and right, respectively. Middle: $H_2$ (0,0) $S$(11) (4.181 μm) and a sum of CO $v = 1$–0 $P$(11)–$P$(18) lines (4.76–4.84 μm). Bottom: map of the 5 μm scattered light continuum for IRAS 16253 and a three-color image combining the 5 μm continuum cavity (red), the narrower $H_2$ outflow cavity (green), and the spatially resolved jet traced by [Fe II] (blue). The circles on the continuum map mark the apertures for the central (red), jet (white), and scattered light (black) spectra with radii of 0″6, 0″3, and 0″3, respectively. The central protostar position from ALMA is marked with a red star. A background star is marked on the continuum map by a red cross.

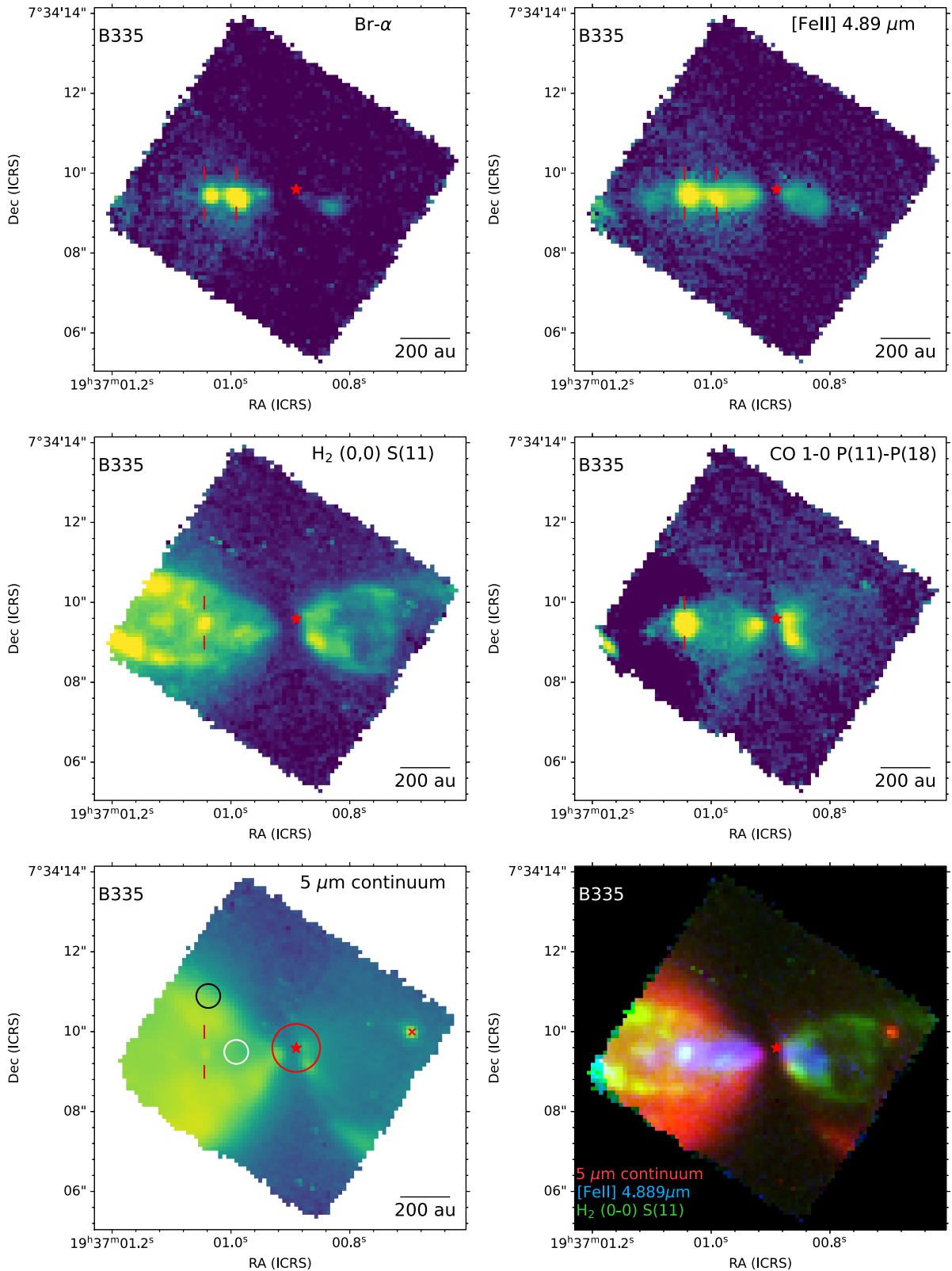

**Figure 3.** Selected line maps for B335, displaying the different morphologies based on outflow tracer. All images are displayed with an arcsinh scale. Top: Brα (4.052 μm) and [Fe II] (4.889 μm) on the left and right, respectively. Middle: H₂ (0,0) S(11) (4.181 μm) and a sum of CO $v = 1$–0 $P(11)$–$P(18)$ lines (4.76–4.84 μm). Bottom: map of the 5 μm scattered light continuum for B335 and a three-color image combining the 5 μm continuum cavity (red), the narrower H₂ outflow cavity (green), and the spatially resolved jet traced by [Fe II] (blue). Red dashes mark prominent knots of shocked emission. The faint knot seen in the continuum map is an artifact of the blending of the CO line wings. The circles on the continuum map mark the apertures for the central (red), jet (white), and scattered light (black) spectra with radii of 0″.6, 0″.3, and 0″.3, respectively. The central protostar position from ALMA is marked with a red star. A background star is marked on the continuum map by a red cross.

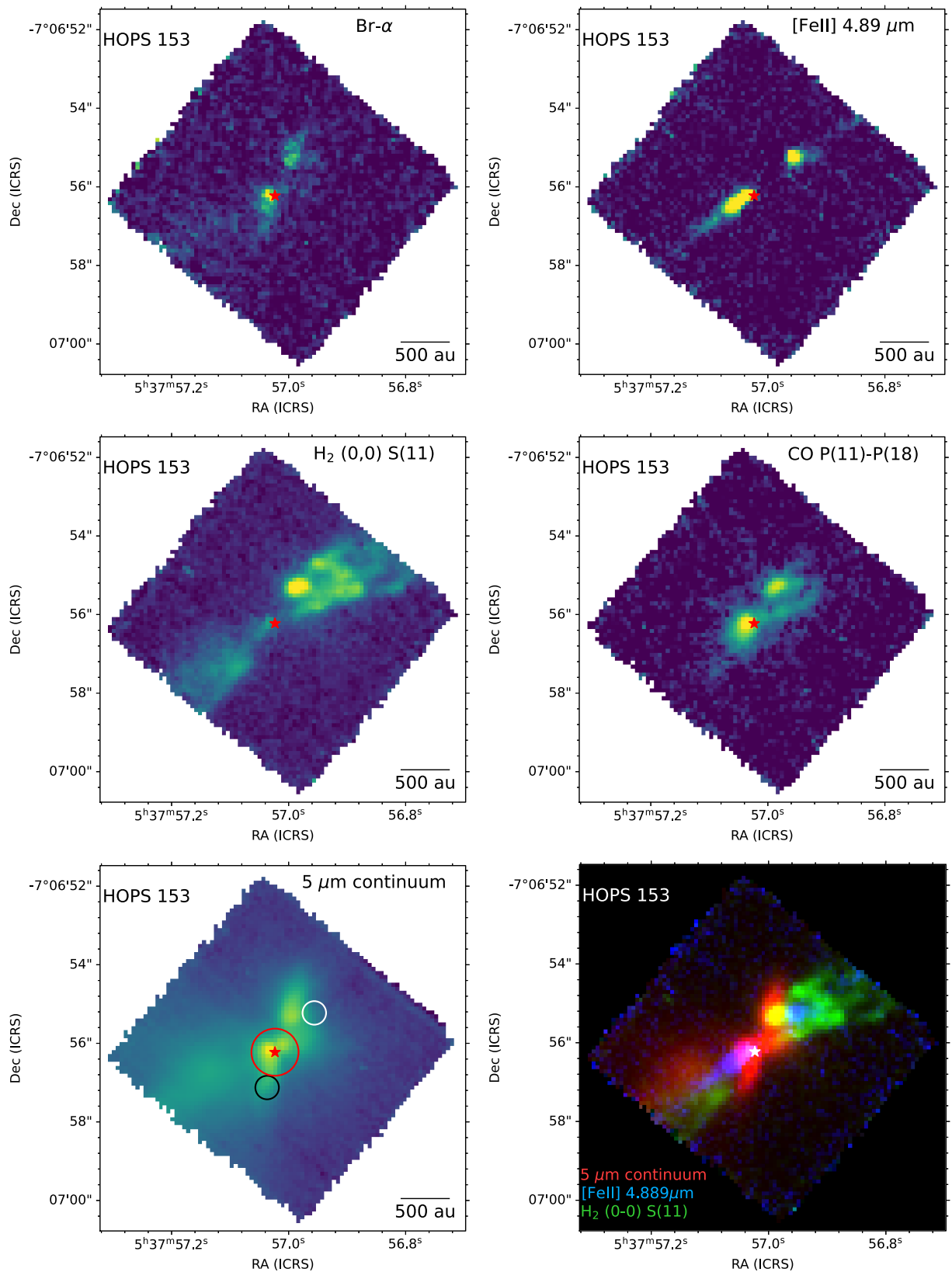

**Figure 4.** Selected line maps for HOPS 153, displaying the different morphologies based on outflow tracer. All images are displayed with an arcsinh scale. Top: Brα (4.052 μm) and [Fe II] (4.889 μm) on the left and right, respectively. Middle: $H_2$ (0,0) S(11) (4.181 μm) and a sum of CO $v = 1–0$ P(11)–P(18) lines (4.76–4.84 μm). Bottom: map of the 5 μm scattered light continuum for HOPS 153 and a three-color image combining the 5 μm continuum cavity (red), the narrower $H_2$ outflow cavity (green), and the spatially resolved jet traced by [Fe II] (blue). The circles on the continuum map mark the apertures for the central (red), jet (white), and scattered light (black) spectra with radii of 0″6, 0″3, and 0″3, respectively. The central protostar position from ALMA is marked with a red star (white for bottom right panel).

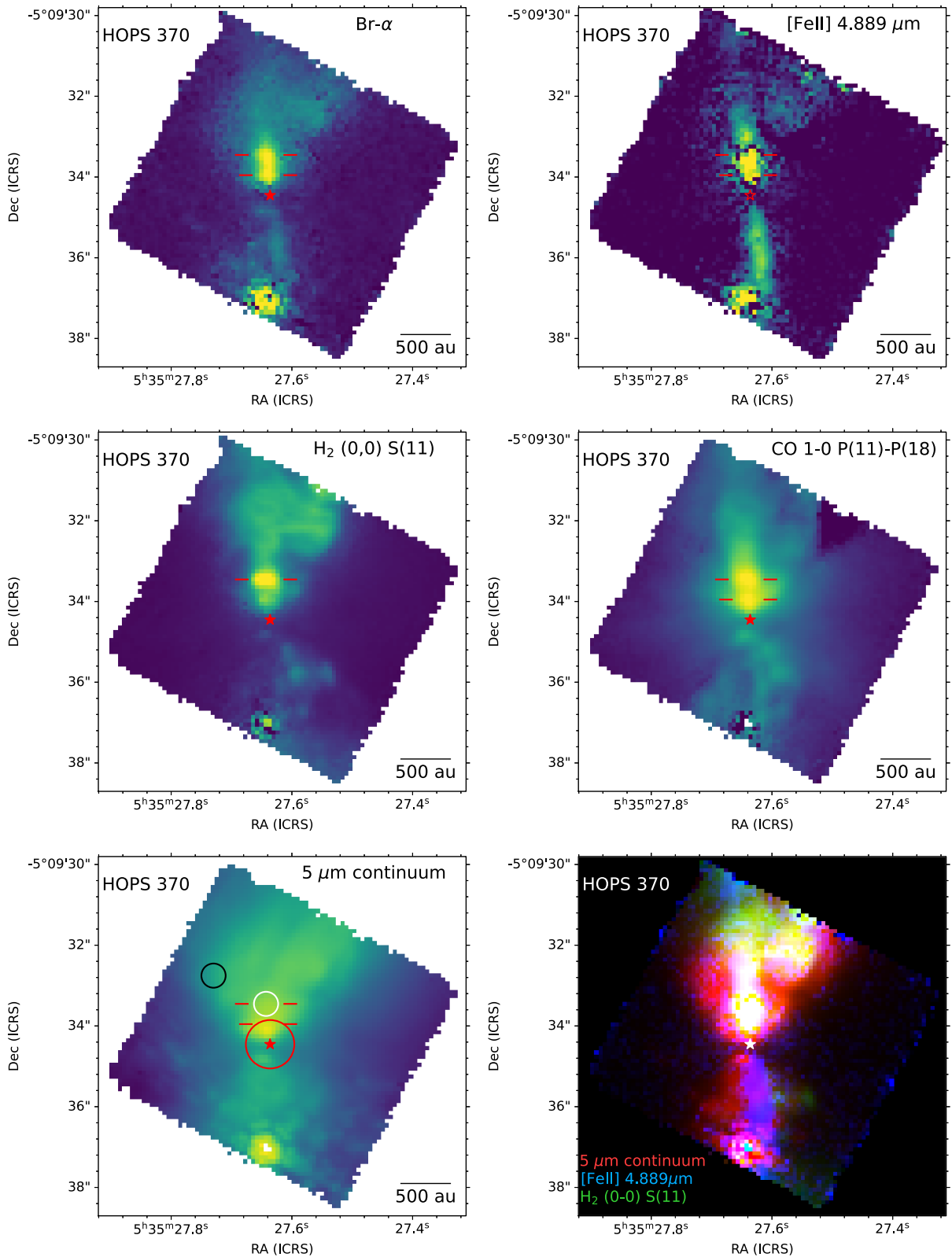

**Figure 5.** Selected line maps for HOPS 370, displaying the different morphologies based on outflow tracer. All images are displayed with an arcsinh scale. Top: Brα (4.052 μm) and [Fe II] (4.115 μm) on the left and right, respectively. The dark pixels across the jet in the [Fe II] image are an artifact of the subtraction of the continuum and CO emission. Middle: H$_2$ (0,0) $S$(11) (4.181 μm) and a sum of CO $v = 1$–0 $P$(11)–$P$(18) lines (4.76–4.84 μm). Bottom: map of the 5 μm scattered light continuum for HOPS 370 and a three-color image combining the 5 μm continuum cavity (red), the narrower H$_2$ outflow cavity (green), and the spatially resolved jet traced by [Fe II] (blue). Red dashes mark prominent knots of shocked emission. The circles on the continuum map mark the apertures for the central (red), jet (white), and scattered light (black) spectra with radii of 0″.6, 0″.3, and 0″.3, respectively. The central protostar position from ALMA is marked with a red star (white for bottom right panel). A saturated background star is visible at the southern edge of the field.

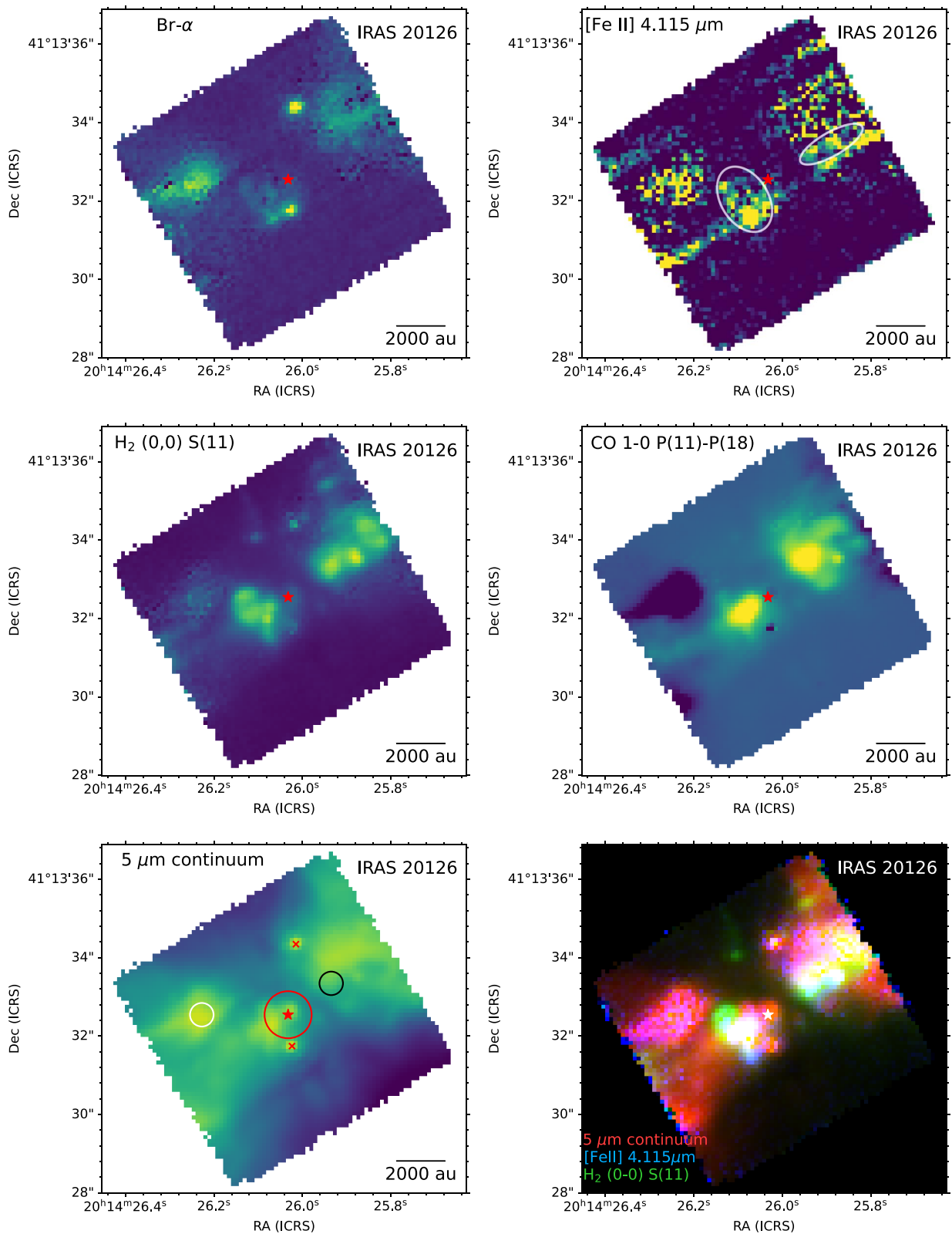

**Figure 6.** Selected line maps for IRAS 20126, displaying the different morphologies based on outflow tracer. All images are displayed with an arcsinh scale. Top: Brα (4.052 μm) and [Fe II] (4.115 μm) on the left and right, respectively. The faded white ellipses in the [Fe II] map contain bona fide emission; emission outside of the ellipses is an artifact of the continuum subtraction. Middle: H₂ (0,0) *S*(11) (4.181 μm) and a sum of CO *v* = 1–0 *P*(11)–*P*(18) lines (4.76–4.84 μm). Bottom: map of the 5 μm scattered light continuum for IRAS 20126 and a three-color image combining the 5 μm continuum cavity (red), the narrower H₂ outflow cavity (green), and the spatially resolved jet traced by [Fe II] (blue). The circles on the continuum map mark the apertures for the central (red), jet (white), and scattered light (black) spectra with radii of 0″6, 0″3, and 0″3, respectively. The two likely companions are marked on the continuum map by red crosses. The central protostar position from ALMA is marked with a red star (white for bottom right panel).

### 3.2.2. B335

B335 provides an example of a low-mass protostar in an isolated globule with a very small disk (radius <10 au). Wide-field Infrared Survey Explorer (WISE)/NEOWISE 4.5 μm photometry of B335 indicates that its luminosity increased by a factor of 5–7 starting between 2010 and 2015 but has now returned to its previous luminosity (Evans et al. 2023). The 33 au resolution NIRSpec image of the inner 1000 au of the B335 protostar shows the bipolar morphology first apparent at much larger scales in Spitzer imaging (see Figure 3 and Stutz et al. 2008). The 5 μm scattered light continuum (bottom left panel) shows the full extent of the cavity. Here the eastern cavity is brighter than the western cavity, indicating that the eastern side is inclined toward us. This is consistent with millimeter observations of CO that indicate that the outflowing gas associated with the eastern cavity is blueshifted relative to the systemic velocity (Yen et al. 2010). In the eastern cavity, a jet containing multiple knots is apparent in [Fe II] (top right panel). A jet is also visible in the western cavity. This jet is more affected by extinction due to the inclination away from the observer but is wider than the eastern jet (Section 3.2.6). There is also faint, extended [Fe II] emission that reaches the edge of the eastern cavity.

The Brα map (top left panel) shows two bright knots coincident with [Fe II] knots, as well as faint emission along the eastern and western jets. The eastern Brα/[Fe II] knot is also apparent in $H_2$ (middle left panel) and the CO *P*-branch (middle right panel). Although this knot also faintly appears in the 5 μm scattered light continuum image, this is an artifact due to the blending of the CO line wings resulting in a slight overestimate of the continuum. There is no significant $H_2$ or CO emission observed toward the western knot. The eastern knot, which is farther downstream than the western knot, is 12 times brighter in [Fe II] emission relative to Brα. This is in contrast to the western knot, in which the [Fe II] emission is roughly equal to that of Brα.

The warm $H_2$ traces shell structures similar to IRAS 16253, with rich internal substructure. Unlike IRAS 16253, the eastern $H_2$ shell of B335, which is slightly inclined toward the observer, is brighter than the emission from the western cavity. Similar to the northern shell of IRAS 16253, the $H_2$ shells are narrower than the full extent of the outflow cavities as traced by the 5 μm continuum, as shown in the color-composite image (bottom right panel). In the eastern cavity, faint $H_2$ emission is seen extending beyond the narrow $H_2$ shell and filling the cavity defined by the continuum. The CO *P*-branch emission (middle right panel) shows a bright, wide-angle morphology that connects to the base of the eastern cavity and follows the $H_2$ emission along the base of the western cavity. A background star is apparent toward the western cavity. A background-subtracted spectrum of this star can be found in Appendix D.

### 3.2.3. HOPS 153

HOPS 153 in the filamentary Orion A cloud is the last low-mass protostar in our sample, with a disk over 10 times larger than the other low-mass protostars in our sample (Table 1). NIRSpec imaged a 2340 au region with 78 au resolution. Although bipolar, HOPS 153 does not show the symmetric hourglass morphology exhibited by the two previous sources (Figure 4). Hubble Space Telescope (HST) and Spitzer imaging indicate that the southeastern cavity is inclined toward us (Habel et al. 2021). The 5 μm scattered light continuum (bottom left panel) traces the bright southern half of the southeastern outflow cavity and the bright northern half of the northwestern outflow cavity, both with "comma"-shaped structures. A fainter peak of emission is located between the two knots, closer to the southeastern knot. Fainter, more diffuse emission partially fills the cavities, particularly in the southeastern cavity.

The [Fe II] jet (top right panel) shows a tightly collimated, bipolar morphology. The northwestern [Fe II] jet is extinguished owing to its inclination of that cavity away from the observer. The [Fe II] jets are bright near the base, but there are faint components that extend much farther. The Brα (top left panel) is concentrated in two peaks that are coincident with the base of the scattered light continuum, offset from the axis of the [Fe II] jet.

In the northwestern jet of HOPS 153, there appears to be one knot observed in [Fe II] emission and another offset from the main jet axis observed in Brα, $H_2$, and CO emission. This second knot is coincident with bright scattered light emission and may be scattered light from more obscured regions near the central protostar. The southeastern jet of HOPS 153 shows two [Fe II] knots in close proximity; there is also Brα and CO emission at this location slightly offset from the [Fe II] knots. As in the northwestern outflow, its proximity to the continuum peaks suggests that these lines are emission from the obscured region seen in scattered light.

Similar to IRAS 16253, the outflow cavity inclined away from the observer, i.e., the northwestern cavity, is brighter in $H_2$ emission. In this cavity, there is a narrow shell-like structure partially covered in the NIRSpec FOV. In contrast to the shells in IRAS 16253 and B335, the HOPS 153 shell has a more complex structure, including a bright $H_2$ knot. As in the case of the previous two protostars, outflow cavities defined by the scattered light are much wider than the emission apparent in the $H_2$ *S*(11) line (bottom right panel). Likewise, the CO *P*-branch emission (middle right panel) in the southeastern cavity is concentrated at the continuum peak at the base of the outflow cavity. In the northwestern cavity, the CO emission is more extended and appears to follow the basic contours of the $H_2$ outflow with a bright knot of emission adjacent to the continuum, $H_2$, and Brα knots.

### 3.2.4. HOPS 370

HOPS 370 is the intermediate-mass protostar in our sample at the Class 0/I boundary and one of the few detected cases of a protostar driving a nonthermal jet (Osorio et al. 2017). At the same distance as HOPS 153, NIRSpec imaged a 2340 au region with 78 au resolution. Over this region, the continuum also exhibits an hourglass morphology, with a bright northern cavity and faint southern cavity indicating that the northern cavity is tilted toward the observer (Figure 5). The 5 μm scattered light continuum (bottom left panel) shows the full extent of the cavity, although the emission is concentrated toward the interior of the cavity with fainter, more diffuse edges. The emission from the cavities has an irregular structure that might be partly due to structures in the foreground. At the southern end of the field, there is a bright pre-main-sequence star/ Class II object seen in the continuum (Megeath et al. 2012), but since it is saturated, we do not extract a background-subtracted spectrum for this star.

The [Fe II] emission (top right panel) traces prominent, bipolar jets. Two knots are present in the northern jet, with the southern knot coincident with bright continuum emission at the base of the northern cavity. The dark pixels across the jet in the image are an artifact of the extraction of the line emission from the continuum and CO forest. The Brα line (top left panel) also shows a clumpy, collimated jet-like structure dominated by two prominent knots in the northern cavity coincident with the [Fe II] knots. The southern jet is faintly visible in Brα.

The warm $H_2$ emission (middle left panel) shows a great amount of substructure. As for the previous protostars, the $H_2$ emission is narrower than the scattered light cavity (bottom right panel). The base of the $H_2$ emission in the northern cavity is jetlike before running into a bright knot of emission coincident with the northern bright knot seen in Brα and [Fe II]; this molecular jet is wider than the ionic jet observed in [Fe II] (Section 3.2.6). The $H_2$ emission past the bright knot has a "mushroom-cloud" morphology, with a narrow stem—this may be the continuation of the jet—connecting to broader emission to the north. This broader $H_2$ emission shows a wispy substructure, with a large lobe extending to the west. This lobe is coincident with faint emission in Brα and [Fe II]. In the southern cavity, there is also a faint, clumpy arc of emission evident in the $H_2$ that is offset to the west of the [Fe II] jet.

In the northern cavity, the CO *P*-branch emission (middle right panel) shows a wide clumpy jet component very similar to that seen in the [Fe II] jet at the base and growing wider farther from the source. The two jet knots detected in [Fe II] and Brα are also seen in the CO *P*-branch. There also appears to be an extended CO emission component that follows the edge of the outflow cavity. Clumpy CO emission is visible in the southern outflow lobe, which is more extended than the jet seen in [Fe II] and coincident with the arc of $H_2$ emission.

### 3.2.5. IRAS 20126

IRAS 20126 is the most massive protostar in our sample, and at a distance of 1.6 kpc (Moscadelli et al. 2011), it is the source we observe with the coarsest linear spatial resolution (320 au) and largest field (9600 au). A molecular outflow originating from this source was first revealed in 2.12 μm $H_2$ and 3.5 mm SiO emission by Cesaroni et al. (1997, 1999), and knots in a jet were observed at subarcsecond resolution by the Very Large Array in several centimeter continuum bands (Hofner et al. 2007). A bipolar structure around the nascent B-type star at 3–5 μm was resolved at much larger scales by Spitzer (Qiu et al. 2008). The IFU maps for IRAS 20126 show the base of that bipolar cavity in the 5 μm scattered light continuum (bottom left panel of Figure 6). The structure of the continuum emission is more irregular than that evident in the lower-mass protostars. Toward the cavities are two likely companions that appear as point sources; the northern one is found in 2 μm *K*-band imaging by Cesaroni et al. (2013), and the southern one is found at mid-IR wavelengths (Sridharan et al. 2005). Both show Brα in emission, suggesting that they are young stellar objects (YSOs; Appendix D). This is not surprising, as most OB-type stars occur in double or higher-order multiple systems (e.g., Duchêne & Kraus 2013). A compact continuum source is visible at the apex of the southeastern cavity and close to the millimeter position of protostar; this source lacks corresponding Brα emission.

The ionic and molecular line emission traces bipolar structures inside the cavities centered on the millimeter protostar location (bottom right panel). The [Fe II] emission (top right panel) traces two compact knots in the northwestern outflow, lacking the continuous, collimated morphology evident in the jets of the lower-mass protostars. The [Fe II] emission in the southeastern outflow is blueshifted and even less collimated than the northwestern outflow, spanning a wide angle with substructures to the north and south. Due to the difficulty in extracting the relatively weak [Fe II] 4.115 μm emission from a sharply rising continuum, the [Fe II] image also shows noisy, extended emission, which is an artifact of the continuum subtraction. The bona fide [Fe II] emission is contained within the faded white ellipses (Figure 2, top right panel); this has been verified through the spectra of the various regions, as shown in Appendix E, and through comparison with the 5.34 μm emission from MIRI (private communication). The Brα emission coming from IRAS 20126 is diffuse except for the two point sources; this is in contrast to the [Fe II] and to the Brα knots and collimated jet components detected toward lower-mass protostars.

The $H_2$ emission (middle left panel) shows an irregular, clumpy bipolar morphology. Structured wide-angle $H_2$ emission in the southeastern cavity is coincident with the southeastern [Fe II] emission. In the northwestern cavity, the $H_2$ shows two knots coincident with the [Fe II] knots, extended emission extending along the knots, and a structure directly ahead of the knots; this structure is coincident with more diffuse emission in the CO *P*-branch and an arc-like feature in the continuum that is also seen in the 2 μm continuum (Cesaroni et al. 2013). Bright, extended CO *P*-branch emission is detected toward the [Fe II] knots and $H_2$ emission. Diffuse emission is also apparent in the CO *P*-branch, which follows structures apparent in the continuum and Brα; these lines may be tracing emission from the outflow or from the central protostar seen in scattered light. A large region of CO *P*-branch absorption is coincident with a region of bright scattered light continuum roughly due east of the protostar position.

### 3.2.6. Measuring Jet Properties

To quantify the properties of the jets and the knots found in the jets, we determine the widths of the jets and positions of the knots. To measure the widths perpendicular to the jet axes, we rotate each [Fe II] image in Figures 2–6 so that the jets are aligned vertically. The rotation angle for each image is listed in Appendix Table 8. We then fit Gaussian or Lorentzian profiles plus a baseline to horizontal cuts across the brightest point in each [Fe II] jet; the resulting FWHMs are in Table 3. Since the measured widths are close to the angular resolution of NIRSpec, we also provide deconvolved jet widths corrected for the angular resolution by subtracting out the FWHM of a point source in quadrature. The FWHM is measured using the

| Jet[a] | Width | Deconv. Width | |
|---|---|---|---|
| | (arcsec) | (arcsec) | (au) |
| IRAS 16253-N | 0.246 ± 0.001 | 0.144 ± 0.001 | 20.2 ± 0.1 |
| B335-E | 0.301 ± 0.001 | 0.225 ± 0.001 | 37.1 ± 0.1 |
| B335-W | 0.662 ± 0.002 | 0.631 ± 0.002 | 104.1 ± 0.3 |
| HOPS 153-NW | 0.287 ± 0.001 | 0.206 ± 0.001 | 80.1 ± 0.1 |
| HOPS 153-SE | 0.277 ± 0.001 | 0.191 ± 0.001 | 74.6 ± 0.1 |
| HOPS 370-N | 0.323 ± 0.001 | 0.261 ± 0.001 | 101.9 ± 0.1 |
| HOPS 370-N H$_2$[b] | 0.479 ± 0.001 | 0.439 ± 0.001 | 171.3 ± 0.1 |
| HOPS 370-S | 0.411 ± 0.001 | 0.364 ± 0.001 | 141.9 ± 0.1 |
| IRAS 20126-NW | 0.241 ± 0.001 | 0.148 ± 0.001 | 236.1 ± 0.3 |

**Notes.**
[a] Jet widths are measured in the [Fe II] lines shown in Figures 2–6 except when marked as H$_2$.
[b] Jet width as measured in the H$_2$ 0–0 $S$(11) line.

point source in the IRAS 16253 image; its value is 0″2 at 4.889 $\mu$m and 0″19 at 4.115 $\mu$m. In HOPS 370, we also measured the width of the northern jet in the H$_2$ 0–0 $S$(11) line; this is the only source where an H$_2$ jet is detected. Converting the angular widths to spatial widths using our adopted distances, we find that the jets range in width from 20 au in IRAS 16253 to 236 au in IRAS 20126.

After identifying knots by eye, their positions along the jet axis are determined by fitting Gaussian or Lorentzian profiles to cuts along the jets, with the choice of function depending on the profile of the knot. We associate knots detected in different tracers if their centers are separated by less than one resolution element (0″2). The projected vertical distance from knot centers in each tracer to the point on the jet axis closest to assumed protostar positions (the position from which the central spectrum is extracted) is listed in Table 4. An example is shown in Appendix F. The size of a knot is calculated using the FWHM of the Gaussian or Lorentzian fit to the knot profile. Also in Table 4 are the corresponding dynamical timescales for the mean of the center knot positions (distance/velocity) in all available tracers assuming ballistic motions, a fiducial jet velocity of 100 km s$^{-1}$ (e.g., Lee 2020; Narang et al. 2024), and the inclinations listed in Table 1. These results are discussed in Section 4. We also tabulate the coordinate positions of the knots in the jets (Appendix F). The knots seen in molecular emission and Br$\alpha$ toward HOPS 153 are strongly associated with the continuum and offset from the [Fe II] jet (Figure 4). These could be from scattered light emission and are not tabulated because they do not appear to be associated with the jet.

### 3.3. Comparing Jet and Outer Cavity Spectra

To further explore the variations in the spectra, we show spectra for each source extracted from an aperture centered on the jet and on a position near the edge of the scattered light cavity in Figure 7. The jet positions are centered on prominent knots of [Fe II] emission or the brightest point in the continuous [Fe II] jet in the case of IRAS 16253. In contrast, the position on the edge of the cavity is relatively free of the bright line emission detected from the jet. The positions are displayed in the 5 $\mu$m continuum images for Figures 2–6.

The jet position is characterized by strong emission lines. Strong [Fe II] emission is found toward IRAS 16353, B335,

and HOPS 153, and weaker [Fe II] emission is associated with HOPS 370 and IRAS 20126. Br$\alpha$ is also detected in IRAS 16253, B335, HOPS 153, and HOPS 370. Finally, due to the choice of the location along the jet, numerous bright H$_2$ lines are apparent. Bright CO emission is also evident, particularly for HOPS 370 and IRAS 20126. The continuum at these positions is from the scattered light at the position of the jet, and the ice features are due to absorption in the foreground; we do not detect continuum emission from the jets. These spectra show the full range of ionic, atomic, and molecular emission lines present in or around the jets.

The spectra at the edges of the cavities are dominated by a mixture of scattered light, tracing the light coming from the disk and protostar, and line emission from shocked gas at or near the cavity walls. The 3–5 $\mu$m continuum emission primarily originates in the inner disk, with perhaps some component from the inner envelope, where dust grains are heated to high temperatures near the sublimation radius (e.g., Muzerolle et al. 2003; D'Alessio et al. 2006; Furlan et al. 2011). Against this continuum, the ice features present toward the central and jet positions are also apparent at the cavity edges. At the cavity edges, the line emission is weaker than toward the central position or the jet. The brightest lines are typically those of H$_2$, indicating that the warm molecular gas extends throughout the cavities. In Appendix C we list the lines detected in the jet and scattered apertures, marked by the colored dashes at the top of the spectra in Figure 7.

B335 is the only protostar for which we detect 4.115 $\mu$m [Fe II] emission at the cavity edge; this emission is much weaker than that at the jet position. Br$\alpha$ is detected in the sources, but bright emission is only found toward HOPS 153. The cavity edge location of HOPS 153 has the highest ratio of Br$\alpha$ line to continuum in the spectra shown in Figures 1 and 7; this may be emission from the central protostar detected in scattered light. The CO spectra are more complex; once again, these may be composed of a combination of scattered light from the inner disk, emission from hot gas in the cavity, and absorption lines from hot gas in the cavity seen in the scattered light from the disks (e.g., Herczeg et al. 2011; Brown et al. 2013). Bright CO emission is detected in IRAS 16253, HOPS 153, and HOPS 370, while B335 shows very weak CO emission. In HOPS 370, the CO has the brightest lines in the spectrum, although a narrower range of the *R*- and *P*-branches are excited, as the high-*J P*-branch CO lines are either very faint or in absorption. In IRAS 20126, the CO from the scattered light position is seen fully in absorption rather than emission. Like the central position, IRAS 20126 again appears distinct from the lower-mass protostars, characterized by a strong continuum with only weak Br$\alpha$ emission and CO absorption. Since these emission and absorption lines follow structures in the continuum scattered light, they may be seen in scattered light from the central protostar.

## 4. Discussion

In this section, we discuss the physical components of the protostellar outflows traced by the atomic, ionic, and molecular lines mapped above. For a detailed analysis of the jet from IRAS 16253, see Narang et al. (2024). For a detailed analysis of the CO fundamental series emission from the protostars, see Rubinstein et al. (2023a).

**Table 4**
Projected Distances, Sizes, and Dynamical Timescales of Knots

| Jet Knot[a] | [Fe II] Dist./Size[b] (au) | Brα Dist./Size (au) | H2 Dist./Size (au) | CO Dist./Size (au) | Time[c] (yr) |
|---|---|---|---|---|---|
| IRAS 16253-N | 428/53 | 416/52 | ... | ... | 22 |
| B335-E E | 362/42 | 349/54 | 376/46 | 375/38 | 18 |
| B335-E W | 227/64 | 226/54 | ... | ... | 15 |
| HOPS 153-NW | 533/120 | ... | ... | ... | 27 |
| HOPS 153-SE E | 247/108 | ... | ... | ... | 12 |
| HOPS 153-SE W | 60/93 | ... | ... | ... | 3 |
| HOPS 370-N N | 341/235 | 345/253 | 397/159 | 385/260 | 18 |
| HOPS 370-N S | 198/170 | 218/146 | 216/161 | 201/172 | 11 |
| IRAS 20126-NW W | 3253/384 | ... | 3221/440 | 3165/699 | 155 |
| IRAS 20126-NW E | 2272/546 | ... | 2258/666 | 2253/937 | 109 |

**Notes.**
[a] Individual knots are designated N(orth), S(outh), E(ast), and W(est) when there are two knots in a jet.
[b] Locations and sizes of knots are given by the center and FWHM of the Gaussian or Lorentzian profile fits to the knot profiles. Projected distances are from the adopted protostar positions to the centers of the knots along the axis of the jet.
[c] Dynamical timescales to the average of the knot centers in each tracer are estimated assuming a velocity of 100 km s$^{-1}$ and the inclinations listed in Table 1.

### 4.1. Collimated Bipolar Jets

Accretion-driven, collimated jets are detected across the full evolutionary span of accreting YSOs from Class 0 protostars to pre-main-sequence stars hosting gas-rich disks, i.e., Class II objects (e.g., Frank et al. 2014; Bally 2016; Ray & Ferreira 2021). Collimated jets are detected in the [Fe II] 1.64 μm line toward a number of protostars, including several Class 0 protostars (e.g., Stapelfeldt et al. 1991; Davis et al. 1994, 2011; Gredel 1994; Reipurth et al. 2000; Nisini et al. 2002; Caratti o Garatti et al. 2006; Antoniucci et al. 2014; Erkal et al. 2021). Detected collimated jets, however, are not ubiquitous (e.g., Santangelo et al. 2015). Toward Class 0 protostars, the fact that observed jets are not more commonly observed may be due to obscuration by the high extinction of the protostellar envelopes (e.g., Habel et al. 2021). Jets from the most embedded protostars are often detected in molecules such as SiO in millimeter interferometric observations (e.g., Cesaroni et al. 1997, 1999; Gueth & Guilloteau 1999; Tychoniec et al. 2019; Lee 2020; Dutta et al. 2023); however, many Class 0 protostars, particularly those with the lowest bolometric luminosities, do not show molecular jets (e.g., Podio et al. 2021). This may be explained by the formation of molecules in jets with high enough mass-loss rates, and therefore densities, to shield molecules forming in warm jet gas (Tabone et al. 2020).

In the wavelength range accessible by NIRSpec, all five IPA protostars are traced consistently in ionized [Fe II] emission. For the four IPA protostars with stellar masses ≲2.5 $M_\odot$, the [Fe II] emission appears in the form of collimated jets and knots. For IRAS 20126, we only see knots of [Fe II] along the previously known H2 jet. Although we are limited by the small size of the sample, this result suggests that collimated jets with internal shock velocities sufficient to ionize and excite Fe (>30 km s$^{-1}$; Hollenbach & McKee 1989) are common during the deeply embedded primary infall and accretion phase of protostars.

As discussed in Section 3, the diameters of all the jets are resolved by NIRSpec, although only marginally for IRAS 16253 (Table 3). For the ≲2.5 $M_\odot$ protostars, the jets have deconvolved widths ranging from 20 to 141 au. This rapid collimation of protostellar jets to widths of 10–100 au achieved within 100 au of the protostars is similar to outflows from Class II protostars as noted by Ray et al. (2007). These widths are also similar to widths found in ground-based IFU measurements of the 1.64 μm [Fe II] line toward Class 0/I protostars by Davis et al. (2011), as well as the jet widths found from 1.64 μm [Fe II] HST images of Class 0/I protostars by Erkal et al. (2021). The widths of the jets can be compared to the diameters of disks measured by ALMA in the dust continuum (Table 1). IRAS 16253, HOPS 153, and HOPS 370 all have [Fe II] jets that are 27%–71% of their disk diameters, while the H2 jet of HOPS 370 has a similar diameter to its disk. In contrast, B335 has jet widths much greater than its disk diameter (<20 au). This variation is consistent with magnetocentrifugal models where the jet is launched with initially wide angles from the inner disk and collimated at larger distances by a helical magnetic field twisted around by the jet (e.g., Blandford & Payne 1982; Ouyed & Pudritz 1999; Anderson et al. 2005; Staff et al. 2015; Pudritz & Ray 2019; Lee 2020). In such models, the radius of the jet, which is launched in the inner region of the disk, is independent of the outer radius of the disk.

The [Fe II] maps show that the jets deviate from bipolar symmetry, with differences in the morphology and potentially excitation conditions of the jet and counterjet. The western jet of B335 has a width of 104 au, much larger than the eastern jet (37 au). The southern jet of HOPS 370 has a width of 142 au, compared to 102 au for the northern jet. These indicate differences in the collimation of the jet on either side of the protostars, perhaps due to differences in the launch radius of the jet material, structure and strength of the magnetic field, or the excitation conditions. The deviation from bipolar symmetry is even more extreme for the [Fe II] emission from the high-mass IRAS 20126, with the southeastern blueshifted outflow lacking any clear collimation over a wide angle, while the northwestern outflow shows relatively compact knots along a single axis. More detailed studies of asymmetries in the jets will be performed in future work using longer-wavelength lines detected with MIRI, which has higher spectral resolution than our NIRSpec observations.

The jets are inconsistently traced in other emission lines. In IRAS 16253, collimated Brα emission is detected extending along the northern [Fe II] jet, as well as emission coincident with the knot at the end of the [Fe II] jet. HOPS 370 also shows collimated Brα emission along the northern jet. Furthermore, knots of Brα emission are found coincident with the [Fe II] jets of the B335 and HOPS 370 protostars. Similar detections of >2 μm H I emission along jets are few. Although atomic hydrogen has been observed in jets in the form of HH objects (e.g., Lizano et al. 1988; Garcia Lopez et al. 2008, 2010; Smith et al. 2010; Reiter et al. 2015; Rubinstein et al. 2023b), in only a few instances has Brγ emission been found along a jet (Beck et al. 2010) or in a jet knot (Davis et al. 2011). To our knowledge, our data show the first collimated protostellar jets detected in the Brα line. In the case of our sample, the bright Brα emission detected in our maps can be mostly attributed to jets, though we cannot discount emission from accretion at the position of the central protostar (e.g., Komarova & Fischer 2020, see Section 4.4).

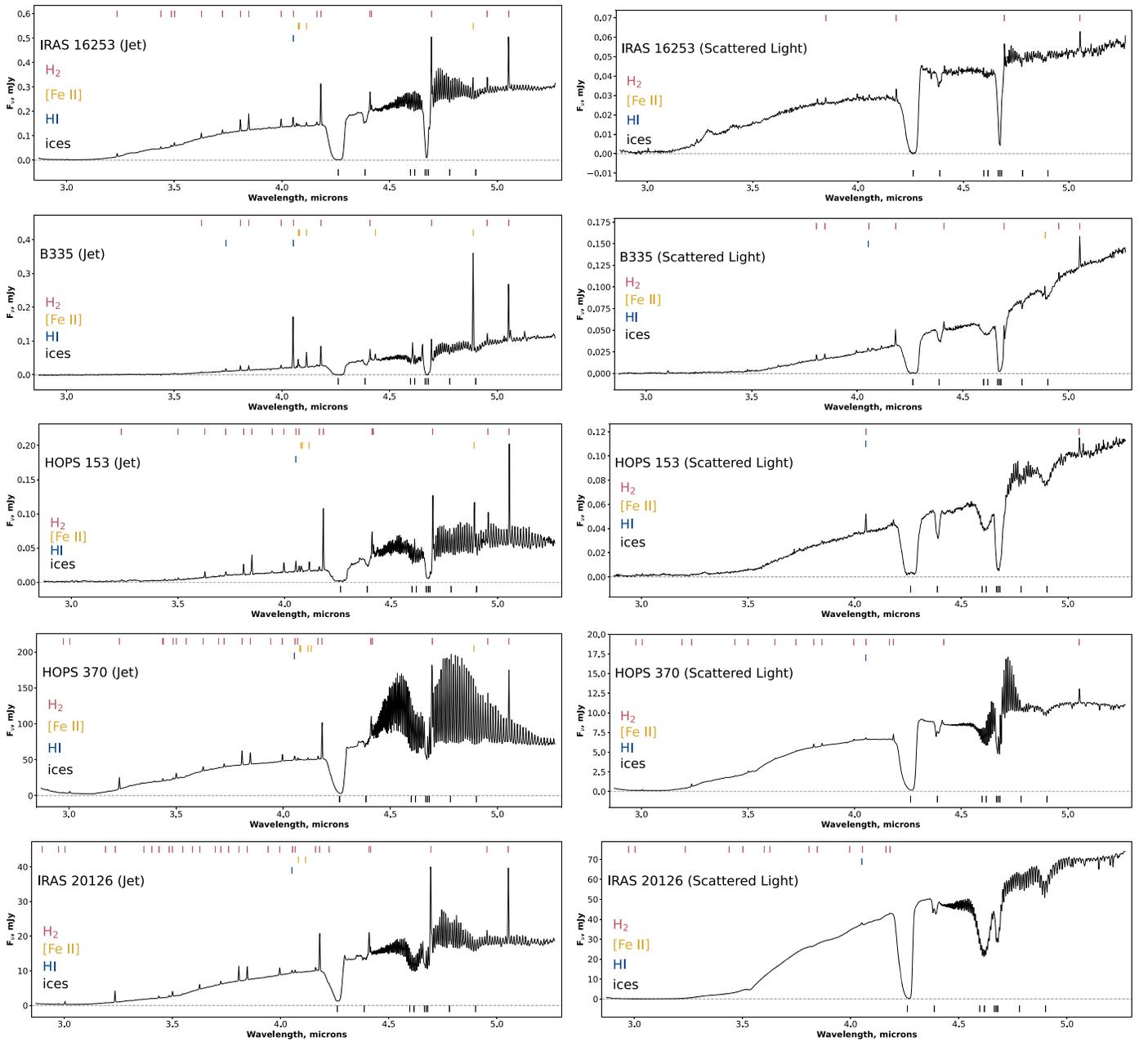

**Figure 7.** Left: NIRSpec IFU spectra from a position located on the jet. Right: spectra from an off-source scattered light position showing light from the central protostar, hot disk, and disk wind. From top to bottom, increasing in protostellar luminosity: IRAS 16253, B335, HOPS 153, HOPS 370, IRAS 20126. The dashes denote the wavelength of H$_2$, atomic and ionic emission lines, and ice absorption features. The spectra are integrated over a 0″.3-radius circular aperture.

Previous observations often detected molecular emission, as traced by CO, SiO, SO, and H$_2$ along the jets of luminous (>1 $L_\odot$) Class 0 protostars (e.g., Nisini et al. 2002, 2015; Tafalla et al. 2004; Dionatos et al. 2009; Lee 2020; Podio et al. 2021;

Hsieh et al. 2023; Ray et al. 2023). In contrast, Class I protostars show jets only in atomic and ionic lines, and low-luminosity (<1 $L_\odot$) Class 0 protostars often do not have detected molecular jets (e.g., Arce et al. 2007; Frank et al. 2014; Podio et al. 2021). In our sample, we observe molecular emission in the jets primarily in the form of concentrated knots, as discussed below. In the northern jet of the intermediate-mass protostar HOPS 370, however, we detect collimated H$_2$ (0–0) $S$(11) and CO fundamental emission. As discussed in Section 3, this molecular jet is wider than the atomic jet. It is also surrounded by fainter H$_2$ emission extending to the edges of the scattered light cavity. The H$_2$ may trace shocks in a slower, collimated molecular component launched at larger radii in the disk than the [Fe II] jet. Alternatively, the H$_2$ emission may be heated in oblique shocks owing to curved shock fronts in the internal working surfaces within the jet. The comparison of the [Fe II] and $H_2$ jets is supportive of the stratified picture of protostellar outflows, with the higher-excitation collimated atomic jet surrounded by a wider molecular jet, which in turn is potentially surrounded by wider-angle molecular wind.

### 4.2. Atomic, Ionic, and Molecular Knots

We define knots as spatially isolated, compact peaks of emission along the jet axis with fluxes much greater than the typical jet emission; the list of knots is given in Table 4. The knots can be traced in atomic, ionic, and molecular lines, even when such tracers are absent in the remainder of the jet. In (sub)millimeter-wave interferometric data, high-velocity molecular knots or "bullets" have been detected in a number of jets (e.g., Bachiller et al. 2000; Hirano et al. 2010; Plunkett et al. 2015; Tychoniec et al. 2019; Podio et al. 2021). Previous observational studies show the presence of symmetrically located knots in the jets and counterjets of protostars; this is evidence that at least some of the knots are associated with variations in the launching of the jet, potentially driven by episodic accretion (e.g., Reipurth 1989; Reipurth & Bally 2001). The knots are primarily embedded within the continuous jets, though they can also be isolated along the jet axis. The knots trace shocks within the jets that likely originate in internal working surfaces, layers of shock-heated gas resulting from higher-velocity material colliding with slower-moving material in the jet (e.g., Schwartz 1983; Raga & Kofman 1992; Reipurth & Heathcote 1992; Suttner et al. 1997; Völker et al. 1999; Nisini et al. 2002). Models and simulations invoking a jet with an oscillating velocity but a constant mass flow rate produce chains of such knots (Raga et al. 1990; Raga & Kofman 1992; Rabena-nahary et al. 2022; Rivera-Ortiz et al. 2023).

For each knot, we have an approximate dynamical time assuming ballistic motions at a fiducial jet velocity of $100 \, \mathrm{km \, s^{-1}}$ (Section 3.2.6). For the $\leqslant 2.5 \, M_\odot$ protostars, this gives timescales of 3–27 yr. In several cases there are two knots in the same jet; the differences in the dynamical times are 3–9 yr between knots. These timescales are much smaller than the ~400 yr timescale between $\geqslant 2$ mag outbursts from Class 0 protostars measured by Zakri et al. (2022); thus, these knots appear to be tracing more frequent variations in the mass flow. If the knots are driven by variations in the accretion rate, the variations must be more rapid and of a smaller amplitude than the IR outbursts reported by Zakri et al. (2022).

While all the knots are detected in [Fe II], the IRAS 16253 knot, both B335 knots, and both HOPS 370 knots are detected in Brα. In these cases, the Brα knots are spatially coincident with the [Fe II] knot to within one instrument pixel, except for the southern HOPS 370 knot, for which the Brα and [Fe II] knot centers are separated by two instrument pixels. This is consistent with the Brα emission originating in the shocks that are also responsible for the [Fe II] emission. In comparison, the Brα knot seen in the northwestern outflow of HOPS 153 is offset by ~0″.5 from the bright [Fe II] knot aligned with the jet and is not included in Table 4. As the Brα emission is coincident with a strong scattered light peak at that position (Figure 4, bottom left panel), we suggest that the Brα emission in this knot is light scattered from the magnetospheric accretion flows onto the protostar.

Knots in B335 and HOPS 370 show strong molecular emission in CO, and two of those knots (B335-E E and HOPS 370-N N) also exhibit bright $H_2$ emission, while the third (HOPS 370-S S) shows comparatively faint $H_2$ emission. Although the heating of the molecular lines also requires shocks, these shocks are offset from the shocks producing the [Fe II] emission. In B335, the eastermost knot appears in CO and $H_2$. In this knot, the molecules are shifted 20 au east (farther downstream) relative to the peak of the [Fe II]/Brα emission. We speculate that this knot is due to a clump of molecular gas launched in a jet with time-varying velocities and gas densities. The [Fe II] emission and Brα emission then originate in a dissociative shock to the west, where the clump is being impacted by faster gas from the jet, while the molecular emission originates in a slower, nondissociative shock on the eastern side, where the clump is overtaking slower-moving gas in the jet (e.g., Neufeld et al. 2009; Harsono et al. 2023; Mohan et al. 2023).

Both knots in the northern jet of HOPS 370 show the presence of molecules; the northern knot shows bright $H_2$ and CO emission, while the southern knot shows bright CO and faint $H_2$ emission. Like the eastern B335 knot, the molecular emission in the northern knot is shifted 50 au north (downstream) of the [Fe II]/Brα knots; this may also show the leading shocks of a molecular clump. In the southern knot of the northern jet, the [Fe II] emission is shifted 18 au south of the $H_2$, yet it is coincident with the CO. The Brα emission here is coincident with the $H_2$. The position of this knot, particularly for the CO emission, may be affected by an additional component of scattered light from the central protostar/disk system. Nevertheless, in all the molecular knots mentioned above, we find a consistent offset with the molecular emission shifted by tens of au downstream of the [Fe II] emission. The knots in the outflows of HOPS 153 do not follow the same trend as B335 and HOPS 370. The Brα knot associated with scattered light in the northwestern outflow of HOPS 153 also shows bright molecular emission from $H_2$ and CO. This indicates that for this particular knot the molecular emission is likely originating from the inner disk region, as is the case for the Brα emission from the same knot.

### 4.3. Scattered Light Cavities and $H_2$ Shells

For all five protostars in our sample, the 5 μm scattered light continuum shows the extents of the outflow cavities. This is expected, as the 4–5 μm continuum traces photons originating from the central protostar and hot inner disk that are scattered off dust grains near the cavity walls. The continuum emission from the two lowest-mass protostars, IRAS 16253 and B335, shows wide-angle scattered light morphologies for the cavities inclined toward the observer; these have a conical or approximately parabolic geometry. The northern cavity of HOPS 370, which is also inclined toward the observer, is also conical but has less clearly defined edges in comparison to the low-mass protostars with a more irregular morphology. In HOPS 153 the scattered light traces primarily the eastern edge of the northwestern cavity and the western edge of the southeastern cavity, with a fainter extended component in the southern cavity. This indicates that HOPS 153 is not found in an axisymmetric core invoked in most collapse models; instead, the cold dense gas may have an elongated, filamentary morphology.

For all of the protostars, $H_2$ emission is detected within the cavities (Figures 2–6). For those with masses $\leqslant 2.5 \, M_\odot$, $H_2$ emission is detected at the cavity walls (Figure 7), demonstrating that warm $H_2$ gas extends to the walls. The bright $H_2$ $S(11)$ line emission, however, typically traces structures inside the cavities. In IRAS 16253, the northeastern cavity shows a bright, partial shell-like structure in the $S(11)$ line that is much narrower than the cavity. This structure extends from the central protostar to the isolated [Fe II] knot in the jet (Figure 2). In the B335 protostar, both cavities contain shells apparent in the $S(11)$ line that extend from the central protostar and are narrower than the cavities; here, the maps do not extend far enough to see whether the apexes of the shells are associated with jet knots (Figure 3). In HOPS 153, the northwestern cavity contains part of a narrow, shell-like structure. The H$_2$ emission in HOPS 370, outside of that from the jet, is more diffuse and fills part of the cavity (Figure 5).

The observed morphology of the H$_2$ emission, as delineated by the $S$(11) line, demonstrates the presence of warm, molecular gas filling the cavities, presumably heated by shocks. In most cavities traced by the continuum, we do not find enhanced emission along the cavity wall in the $S$(11) line, as would be expected for shocks owing to a wide-angle wind colliding with the cavity walls. Instead, we often find narrow shells of H$_2$ that appear similar to the bow shocks and wakes created by internal working surfaces within jets (Tafalla et al. 2017; Tabone et al. 2018; Rabenanahary et al. 2022; Rivera-Ortiz et al. 2023). The coincidence of the apex of the IRAS 16253 shell with the [Fe II]/Brα knot supports this picture. In these models, the cavity may be filled by molecular gas from a disk wind (Tabone et al. 2018) or by gas entrained from the surrounding envelope (Rabenanahary et al. 2022). In this picture, there may be still be lower-velocity shocks beyond the wakes and at the cavity walls.

Only in the southwestern cavity of IRAS 16253 do we see bright $S$(11) emission along the inner cavity walls (middle left panel of Figure 2). The implied shocks along the southwestern inner cavity wall may be due to a wide-angle wind or gas accelerated by the bow shocks colliding with the cavity walls. Thus, it is not clear from the current analysis whether wide-angle winds are required by the current observations.

As an alternative to fast, collimated jets accompanied by slower, wide-angle disk winds, multiple cavities similar to the observed H$_2$ shells are formed in $X$-wind models, where a fast wind collides with a layer of shocked wind gas filling the remainder of the cavity (Shang et al. 2020). A test of these models will be to determine whether emission at the walls of the shells is consistent with the higher-velocity shocks expected from fast $X$-winds or lower-velocity shocks of disk winds.

### 4.4. H I Emission from the Central Protostars

Previous NIR IFU studies of Class I protostars and Class II objects show that the Brγ emission is dominated by compact/ unresolved emission at the central star/protostar (Beck et al. 2010; Davis et al. 2011). Although this central emission is often attributed to magnetospheric accretion (e.g., Muzerolle et al. 1998; Hartmann et al. 2016; Alcalá et al. 2017), interferometric studies of the compact emission suggest that while hydrogen emission lines closest to the protostar originate from magnetospheric accretion, beyond 0.1 au the hydrogen emission may trace the compact inner regions of disk winds (e.g., Kraus et al. 2008; Eisner et al. 2010). Furthermore, JWST/NIRSpec observations of the TMC-1A protostar show NIR H I line profiles suggestive of an origin in a wind (Harsono et al. 2023).

In our sample, the central protostars are too deeply embedded to detect the Brα lines from accretion flows or the inner regions of a wind. The only possible exception is IRAS 16253, where it may be blended with emission from the jet (Figure 2, Section 3.2.6). For these deeply embedded sources, Brα emission from the central protostars may only be detected in the scattered light. Indeed, TMC-1A exhibits spatially extended Paα emission that follows the edge of the outflow cavity and may be scattered light from the central protostar (Harsono et al. 2023).

Brα emission is detected along the edges of the cavities in the spectra displayed in Figure 7. Only in the case of B335 is faint [Fe II] emission also detected at these positions in the NIRSpec data, and there is no other evidence for dissociative shocks at these locations. At these positions, the Brα lines may trace scattered light from the central protostars (Komarova & Fischer 2020). The case for such scattered light is particularly strong for HOPS 153, where relatively bright Brα emission is concentrated toward a scattered light peak (Figures 4 and 7). Further work is needed to establish the detection of scattered light, including studies of the shocks along cavity walls and radiative transfer modeling of the scattered light emission.

### 4.5. Extended CO Emission in the Cavities

Outside of the jets, there is a wider-angle component to the CO emission (Figures 2–6). Ground-based, high spectral resolution spectroscopy of Class I protostars indicates the presence of multiple components in the CO fundamental. These components, which can be distinguished in the line profiles, include emission from a disk, emission from a wind, and absorption by a wind (Herczeg et al. 2011). The number of sources with absorption decreased from 100% to 60% as they dispersed their envelopes and evolved from Class I protostars to Class II objects (Brown et al. 2013). This suggests that some of the absorption in protostars may come from hot gas in shocks along the cavity walls.

The spatially extended CO emission in our Class 0 protostars likely also has multiple components, although at the lower spectral resolution of NIRSpec they cannot be separated by their line profiles. The spatially resolved emission may arise in the shock-heated cavity gas also detected in H$_2$; the origin of some CO emission in outflows is demonstrated by the detection of CO in knots and bow shocks (Section 4.2, Ray et al. 2023). In addition, emission from the inner disk or from a disk wind may be seen in scattered light and thus extend along the cavities (Rubinstein et al. 2023a).

In two protostars, HOPS 370 and IRAS 20126, CO is also observed in absorption against the scattered light continuum. The cavity edge of HOPS 370 shows the higher-$J$ CO transitions in absorption (Figure 7). Toward the central position in IRAS 20126, the higher-$J$ CO transitions are also seen in absorption (Figure 1). Furthermore, in IRAS 20126, all the CO lines are seen in absorption toward a structure in the southeastern cavity (Figures 6 and 7).

The absorption lines can arise as photons from the central disk that travel to the cavity walls and then are scattered. In this case, absorption of the photons can occur in a disk wind, the warm molecular gas filling the cavity, and/or in shocked gas near the cavity edges. If the rotational temperature of the gas is lower than that of the continuum emission from the disk, the corresponding lines will be seen in absorption. Alternatively, for a rapidly accreting protostar, the absorption may occur in the disk; for example, the CO fundamental lines are detected in absorption toward the disk of the outbursting YSO FU Ori (Zhu et al. 2009). The spectrum of the disk with its absorption lines would then be apparent in the spatially extended scattered light in the cavities.

Thus, when there is a mixture of absorption and emission, it may be scattered emission lines from the central disk and a mixture of direct emission and scattered absorption lines from a wind or warm gas in the cavity, or scattered absorption lines from the central disk and direct emission from a wind or warm gas in the cavity. When absorption is present, a given transition will appear in emission or absorption depending on which component dominates. In HOPS 370 and the center of IRAS 20126, emission from a lower rotational temperature gas dominates the lower rotational levels, while absorption from a higher rotational temperature gas dominates the upper levels. In the cavity of IRAS 20126, absorption dominates all the observed rotational levels. In the case of B335, where the CO emission is strong in the inner regions of the cavities and weak near the cavity edges (Figures 3 and 7), the absorption and emission may come close to canceling each other out. Future work will further analyze the CO emission by its rotational diagrams and velocity structure.

### 4.6. Comparing Low-mass Protostars to IRAS 20126

As the highest-mass and most distant protostar in our sample, IRAS 20126 presents similarities to and also significant differences from the lower-mass protostars. Like the lower-mass protostars, we see the inner region of the outflow cavities first observed at lower resolution in Spitzer imaging (Qiu et al. 2008); this appears to be emission from hot dust in the inner regions of the protostars scattered off the walls of the outflow cavities. The cavities are seen on much larger scales than those of the lower-mass stars and exhibit more irregular morphologies. This may be due to structure in the foreground extinction or structural inhomogeneities in the comparatively large cavities. Furthermore, there are two additional YSOs coincident with the cavities that may contribute to the irregular morphology.

Instead of the collimated jets seen in [Fe II] toward low-mass protostars, the [Fe II] emission from IRAS 20126 is weak relative to the continuum and consists of two relatively compact knots along a single axis in the northwest and a wider-angle outflow with two substructures in the southeast. The [Fe II] emission in the northwestern knots may trace parts of a collimated jet similar to those found in the $\leqslant 2.5\,M_\odot$ protostars (Figure 6). In contrast to the lower-mass sources, however, the [Fe II] knots in the IRAS 20126 jet are distant (384–546 au) and have much larger dynamical timescales on the order of a century (109–155 yr), assuming a 100 km s$^{-1}$ velocity (Tables 3 and 4). A 46 yr timescale is calculated between the two knots in the northwestern jet. Given the source distance of 1600 pc, the spatial resolution of the observations (300 au) limits the linear sizes we can measure for this source. Since jet shocks accumulate and weaken rapidly with distance (e.g., Völker et al. 1999), only the strongest and largest shocks should be visible at large scales. It is therefore unclear whether the longer timescales are due to limits in spatial resolution, the accumulation of smaller knots, a faster jet velocity than the assumed 100 km s$^{-1}$, or slower timescale variations in the outflow of this massive protostar leading to more spread-out internal working surfaces.

In contrast to the lower-mass protostars, the bright, structurally complex molecular emission toward IRAS 20126 is found along the jet axis. Spatially extended knots of H$_2$ and CO emission are concentrated along the northwestern jet, coincident with the knots of [Fe II] emission. The H$_2$ emission and CO emission also show wider-angle structures that extend away from the axis of the jet, although the molecular emission is still much narrower than the outflow cavity traced by the scattered light. The implied poor collimation of the jet and the wide outflow cavities may be connected to the extreme precession of the outflow observed on larger spatial scales (e.g., Shepherd et al. 2000; Lebrón et al. 2006; Caratti o Garatti et al. 2008; Cesaroni et al. 2014).

A potential difference between low- and high-mass protostars is the presence of extreme-UV radiation from the hotter, massive protostars. In this case, the high-mass protostars would show H I recombination line emission due to photoionization. The brightest possible potential recombination line that we detect is Brα. The emission from this line is faint and diffuse in IRAS 20126, in sharp contrast to its detection in collimated jets and concentrated knots for the lower-mass protostars. There are no clear knots (excluding the two associated YSOs); the observed emission is likely seen in scattered light from the central protostar. It is not clear whether the central emission is recombination line emission from a photoionized gas or collisionally excited emission from a hot gas. Thus, we find no clear evidence of photoionization by the central protostar.

## 5. Conclusions

We present the first look at NIRSpec IFU data from 2.9 to 5.3 μm for five protostars as part of the IPA JWST Cycle 1 program 1802. The targets have moderate to high disk inclinations, luminosities ranging from 0.2 to 10,000 $L_\odot$, and central protostellar masses of 0.12–12 $M_\odot$. The targets are in their primary accretion phase, and their spectra are inaccessible to NIR observations at 1–2.5 μm. These IFU maps probe the innermost region of protostellar outflows at >3 μm with 0″.2 spatial resolution and $R \sim 1000$ spectral resolution. The spatial resolutions correspond to physical sizes of 30–320 au, and the maps cover fields 840–9000 au in width. These data demonstrate the ability of JWST to reveal critical features of protostellar outflows across logarithmically sampled ranges in protostellar mass and luminosity. Our main results and conclusions are as follows:

1. We detect a number of ubiquitous spectral features across orders of magnitude in protostellar mass and luminosity. The dominant features are a rising continuum at longer wavelengths, deep ice absorption, emission in lines of Brα and [Fe II], and bright molecular emission detected in lines of H$_2$. We detect the CO fundamental series in emission and, for the more massive sources, also in absorption.

2. The 5 μm continuum maps show outflow cavities illuminated by scattered light from the central protostar in all five objects, consistent with bipolar cavities viewed at moderate to high disk inclinations. The cavities of the two lowest-mass protostars show conical/parabolic morphologies, while the other low-mass protostar shows incomplete cavity walls. The cavities of the intermediate- and high-mass protostars exhibit more irregular morphologies, some of which may be due to foreground structures.

3. All of the <2.5 $M_\odot$ protostars show collimated jets in [Fe II] emission bisecting the cavities, while the most massive protostar shows knots in [Fe II]. Although the sample is small, it suggests that in the inner regions of protostars [Fe II] from dissociative shocks is the most reliable tracer of protostellar jets even for young Class 0 protostars.

4. The collimated [Fe II] jets have diameters of 20–142 au; the diameters can differ between a jet and its counterjet. The diameters are independent of the measured disk diameters, which is consistent with magnetocentrifugal models of jet acceleration.

5. Two jets are also apparent in Brα (IRAS 16253 and HOPS 370), and one is apparent in the H$_2$ 0–0 $S$(11) line (HOPS 370). In the latter case, the H$_2$ jet is wider than the [Fe II] jet. To our knowledge, the Brα emission is the first detection of this line in collimated protostellar jets.

6. Compact knots are detected in [Fe II] in all the jets, all but one of which show Brα emission; these may trace shocks from fluctuating velocities in the jets. Three knots also show molecular emission in CO and $H_2$, shifted tens of au downstream from the [Fe II]/Brα emission. These may trace shocks on either side of clumps of molecular gas launched in the fluctuating jets.

7. Assuming a typical jet velocity of $100 \, km \, s^{-1}$, the dynamical timescales of the knots for the $<2.5 \, M_\odot$ protostars are 3–27 yr, while those of the massive source are 109–155 yr. The time between knots in the same jet varies from 3 to 9 yr for the lower-mass protostars, with 46 yr between the knots in the northwestern jet of the highest-mass protostar. This is much shorter than the time between IR outbursts observed toward Class 0 protostars (400 yr). If the knots are produced by time-variable accretion, the variations must be shorter in timescale and smaller in amplitude.

8. Faint $H_2$ emission for the $<2.5 \, M_\odot$ protostars extends to the cavity walls, while bright $H_2$ 0–0 $S(11)$ emission is detected along the walls of only one outflow cavity from IRAS 16253. Instead, we more commonly find the $S(11)$ emission tracing narrow shell structures within the cavities. These maps show that the cavities are filled with warm molecular gas that is likely heated by shocks driven by disk winds and/or by interactions with bow shocks created by internal working surfaces in the central, collimated jets.

9. Brα emission from the central protostars may be primarily apparent in scattered light along the cavities. Efforts to detect accretion flows or inner winds in this line require further modeling to distinguish between scattered light from the central source and line emission from shocks in the cavities.

10. Extended emission in the CO fundamental is detected toward all five protostars, and HOPS 370 and IRAS 20126 have positions that show some or all of the CO transitions in absorption against the scattered light. The observed emission/absorption likely originates in a combination of winds and scattered light from disks.

11. The most massive protostar shows some similarities to the lower-mass protostars, including outflow cavities detected in scattered light and knots of [Fe II] emission along a jet in the cavity center. It also shows some differences from lower-mass protostars, including a very strong deviation from bipolar symmetry; bright, complex structures in the molecular line emission surrounding the jet; and the presence of multiple YSOs coincident within the cavity. The Brα emission from this massive protostar appears to be observed primarily in scattered light from the central source.

## Acknowledgments

This work is based on observations made with the NASA/ESA/CSA James Webb Space Telescope. The data were obtained from the Mikulski Archive for Space Telescopes at the Space Telescope Science Institute, which is operated by the Association of Universities for Research in Astronomy, Inc., under NASA contract NAS 5-03127 for JWST. These observations are associated with program #1802. All the JWST data used in this paper can be found in MAST, doi:10.17909/3kky-t040. Support for S.F., A.E.R., S.T.M., R.G., W.F., J.G., J.J.T., and D.W. in program #1802 was provided by NASA through a grant from the Space Telescope Science Institute, which is operated by the Association of Universities for Research in Astronomy, Inc., under NASA contract NAS 5-03127.

A.C.G. acknowledges support from PRIN-MUR 2022 20228JPA3A "The path to star and planet formation in the JWST era (PATH)" and from INAF-GoG 2022 "NIR-dark Accretion Outbursts in Massive Young stellar objects (NAOMY)" and Large Grant INAF 2022 "YSOs Outflows, Disks and Accretion: toward a global framework for the evolution of planet forming systems (YODA)." E.v.D. is supported by EU A-ERC grant 101019751 MOLDISK. N.J.E. thanks the University of Texas at Austin for research support. A.S. gratefully acknowledges support by the Fondecyt Regular (project code 1220610) and ANID BASAL project FB210003. G.A. and M.O. acknowledge financial support from grants PID2020-114461GB-I00 and CEX2021-001131- S, funded by MCIN/AEI/10.13039/501100011033. P.N. is supported by the Dutch Research Council (NWO) grant 618.000.001. Y.-L.Y. acknowledges support from Grant-in-Aid from the Ministry of Education, Culture, Sports, Science, and Technology of Japan (20H05845, 20H05844, 22K20389), and a pioneering project in RIKEN (Evolution of Matter in the Universe).

We gratefully acknowledge valuable conversations with Dr. Sylvie Cabrit and Dr. Tom Ray.

The University of Toledo acknowledges that the region of Ohio in which the University sits is the ancestral homelands of the Seneca, Erie, and Odawa, as well as places of trade for Indigenous peoples, including the Anishinaabe (Ojibwa, Pottawatomi), Kilatika, Lenape, Kaskaskia, Kickapoo, Miami, Munsee, Peoria, Piankashaw, Shawnee, Wea, and Wyandot. As a steward of public lands, it is our responsibility to understand the history of the land, the peoples who came before us, and their continuing ties to this place. We thank them for their strength and resilience in protecting this land and aspire to uphold our responsibilities according to their example.

## Appendix A
## Custom Masking Process for Data Reduction

The custom masking process consists of several passes to achieve an aggregate performance that we found satisfactory, though imperfect. For our first pass, we download from MAST the Stage 2 raw dark exposures with the same readout pattern and closest in time to our science observations. We perform a statistical analysis of those images, flagging pixels that have count rates significantly larger or smaller than the expected range for the dark signal. We use this to generate a mask, which we then apply to the science data. Many of the remaining bad pixels we observed were generally bright and spatially isolated. For use in the following four passes to remove these pixels, we compute the median image pixel-wise among all 16 Stage 2 "cal" images (i.e., four dithers at four positions) for each target ("med"), and we then compute the median absolute deviation from that median image for each pixel position ("MAD"). These represent outlier-resistant estimates of the typical mean and standard deviation, respectively, for each pixel in the detector for a target's observations. In all passes below, we found that two-level customization of masking thresholds based on source flux performed better than uniform thresholds. We treated IRAS 16253 and B335 as "dim" and the rest as "bright."

Typical behavior includes a bulk correlation between median and MAD pixel values in all observations. Thus, many outliers are identified because they deviate from this locus. Our second pass masks those pixels where the MAD is considerably below that locus:

1. Dim: (MAD < med/1.1–5) OR (med <−5) OR (MAD < 1 AND med > 2)
2. Bright: (MAD < (med+1)/3 AND med > 5) OR (MAD < (med-2)/1.5 AND med > 2 AND med <=5)

The third pass masks those pixels where the MAD is considerably above that locus:

1. Dim: MAD > med/1.1+5
2. Bright: (MAD > med/0.9+5) OR (MAD > $10^6$) OR (med > $10^6$)

Our fourth pass compares med to a boxcar-median-smoothed version of itself using a box size of 5 pixels ("medbox5"). Pixel positions are masked where the med−medbox5 difference is larger than 2.5 or 9 times (bright and dim, respectively) the medbox5 pixel value at that location.

Finally, for the fifth pass, we flag and remove the outermost top and bottom rows of pixels of each IFU slice spectrum. To locate these pixels in the stage 2 "cal" images, we compute a boxcar-median-smoothed version of med using a box size of 10 pixels ("medbox10") and then compute the row-indexed gradient of medbox10 in the upward and downward directions separately. Then, we mask any pixel position where either gradient image is positive. We apply this pass since many of these pixels are unreliable and the other filtering methods are prone to miss them. This pass is the most destructive to potentially good data, however, and thus further improvement to make this pass more selective is desirable. For our purposes, however, this step produces a net improvement in data quality with a limited loss of field of view.

## Appendix B
## Astrometry

In order to perform valid measurements across the NIRSpec IFU and MIRI MRS IFU data sets, as well as to place extant data such as those from ALMA in consistent context with them, we require precise astrometric calibration of each data set. To achieve this, we perform a three-step process that ultimately aligns both NIRSpec and MIRI data for each target to each other and to the ICRS standard through Gaia EDR3 (Gaia Collaboration et al. 2016, 2023). First, we match and merge SESNA catalog products (Gutermuth et al. 2019) with Gaia EDR3 counterparts in a square field of view 10 times larger than the MIRI imaging field of view, and then we apply the mean offset to the SESNA positions to force agreement with Gaia. These offsets are typically less than 0″.1, with uncertainties of ∼3 mas. Second, we use PhotVis (Gutermuth et al. 2008) to identify point sources in the MIRI 7.7 μm bonus imaging obtained in parallel with the MIRI MRS IFU observations of each target. We match the catalog to Gaia-calibrated SESNA counterparts and apply the resultant mean offsets to the MIRI MRS data. Some targets are too extinguished for any Gaia sources to be detected in the MIRI imaging. Third, we perform direct feature matching between the NIRSpec IFU data and MIRI MRS IFU ch1-short data that have substantial spectral overlap (4.90–5.26 μm) in order to find the NIRSpec offset to make it match the Gaia-calibrated MIRI data.

In this final step, we select an emission feature in each field to track on that is nominally continuum emission dominated, unresolved, or centrally concentrated in structure and reasonably isolated from other features. To mitigate the influence of narrow-line features, residual hot spaxels, and other artifacts, we first median-smooth the flux and uncertainty cubes with a 5-pixel window along the spectral axis. This is the main flux cube we use for fitting below. We also use the smoothed flux cube to compute an alternative uncertainty cube. We subtract the median-smoothed flux cube from the unaltered original, take the absolute value for all data in the result, and then median-smooth the result as before. The final result is a local median absolute deviation from the median estimate of the uncertainty for each spaxel.

From these cubes, we compute the mean flux image (and separate propagated error images using each of the uncertainty cube products) over the shared spectral window range along the spectral axis to make a single flux image with two uncertainty maps. Then, we perform a Gaussian fit to each selected feature using mpfit in IDL (Markwardt 2009) within five fitting window sizes (0″.6–1″.4 in 0″.2 steps) all centered on the selected target spaxel position (generally the peak spaxel position). For each window size, we fit twice, once with each uncertainty image as the spaxel weight for the fit. We aggregate the results of the 10 different fits and compute the mean fit uncertainty and the rms of the fit samples, rejecting any obvious outliers. Outliers are defined as being >10$\sigma_{MAD}$, the median absolute deviation from the median of the R.A. and decl. offsets, computed independently by target and axis. The rms value is substantially higher than the fit uncertainty in all cases. Thus, we adopt the former as our reported uncertainty for all measurements. Patapis et al. (2024) estimated a 0″.01 uncertainty in the MIRI astrometry from distortion and 0″.03 reproducibility uncertainty due to the optics of MIRI, which we add to our calculated uncertainties in quadrature. The resulting astrometric offsets applied to each data set and their associated uncertainties, all referenced to Gaia EDR3 through the boot-strapping process described above, are listed in Table 5.

**Table 5**
Applied Astrometric Offsets in Arcseconds

| TargetName-Axis | Gaia-SESNA | σ(Gaia-SESNA) | Gaia-MIRI | σ(Gaia-MIRI) | Gaia-NIRSpec | σ(Gaia-NIRSpec) |
|---|---|---|---|---|---|---|
| IRAS 16253-R.A. | 0.076 | 0.003 | −0.20 | 0.050 | 0.17 | 0.050 |
| IRAS 16253-decl. | −0.031 | 0.003 | −0.53 | 0.040 | 0.14 | 0.045 |
| B335-R.A. | 0.012 | 0.002 | −0.08 | 0.038 | 0.60 | 0.040 |
| B335-decl. | 0.023 | 0.002 | −0.10 | 0.037 | 0.16 | 0.051 |
| HOPS 153-R.A. | 0.032 | 0.004 | 0.17 | 0.040 | −0.20 | 0.068 |
| HOPS 153-decl. | 0.046 | 0.004 | 0.10 | 0.055 | 0.17 | 0.057 |
| HOPS 370-R.A. | 0.020 | 0.003 | 0.15 | 0.047 | −0.24 | 0.047 |
| HOPS 370-decl. | 0.006 | 0.003 | −0.08 | 0.053 | 0.09 | 0.051 |
| IRAS 20126-R.A. | −0.037 | 0.002 | 0.06 | 0.036 | 0.11 | 0.037 |
| IRAS 20126-decl. | −0.005 | 0.001 | −0.04 | 0.039 | −0.14 | 0.040 |

## Appendix C
## Prominent Lines in the Spectra

In Table 6 we list the prominent emission lines in the spectra of Figures 1 and 7, as well as the apertures in which they are detected. The apertures for each source are displayed in the 5 μm continuum images (bottom left panel) of Figures 2–6. For the lines with wavelengths shorter than 4.3 μm that are free of ice absorption features and the CO forest, we report lines with

$\geqslant 5\sigma$ detections in integrated line flux using a simple continuum subtraction, confirmed by visual inspection. In the case of overlapping lines, we mark both as detections if the combined flux of the blended line passes the $\geqslant 5\sigma$ threshold. The ice absorption and CO forest add additional uncertainties in determining the fluxes of overlapping lines with wavelengths longer than 4.3 μm. For these lines we report only those that show clear emission above the level of the nearby CO lines.

**Table 6**
Prominent Emission Lines Observed in the Spectra of Figures 1 and 7

| Species | Wavelength (μm) | IRAS 16253 | B335 | HOPS 153 | HOPS 370 | IRAS 20126 |
|---|---|---|---|---|---|---|
| H$_2$ 1–0 $Q$(16) | 2.898328 | ⋯ | ⋯ | ⋯ | ⋯ | J |
| H$_2$ 2–1 $O$(3) | 2.974064 | ⋯ | ⋯ | ⋯ | C, J, S | C, J, S |
| H$_2$ 1–0 $O$(4) | 3.003868 | C | ⋯ | C | C, J, S | C, J, S |
| H$_2$ 2–1 $O$(4) | 3.189811 | ⋯ | ⋯ | ⋯ | C,S | C, J |
| H$_2$ 1–0 $O$(5) | 3.234988 | C, J | ⋯ | J | C, J, S | C, J, S |
| H$_2$ 1–0 $Q$(21) | 3.298175 | ⋯ | ⋯ | C | ⋯ | ⋯ |
| H$_2$ 0–0 $S$(21) | 3.368711 | ⋯ | ⋯ | ⋯ | ⋯ | J |
| H$_2$ 0–0 $S$(19) | 3.404163 | ⋯ | ⋯ | ⋯ | ⋯ | C, J |
| H$_2$ 2–1 $O$(5) | 3.437867 | C, J | ⋯ | ⋯ | C, J, S | C, J, S |
| H$_2$ 0–0 $S$(18) | 3.438575 | C, J | ⋯ | ⋯ | C, J, S | C, J, S |
| H$_2$ 0–0 $S$(17) | 3.485803 | C, J | ⋯ | ⋯ | C, J | C, J |
| H$_2$ 1–0 $O$(6) | 3.500809 | C, J | ⋯ | J | C, J, S | C, J, S |
| H$_2$ 0–0 $S$(16) | 3.547507 | ⋯ | ⋯ | ⋯ | J | C, J |
| H$_2$ 1–1 $S$(20) | 3.600245 | ⋯ | ⋯ | ⋯ | ⋯ | S |
| H$_2$ 0–0 $S$(15) | 3.626166 | C, J | J | C, J | C, J, S | C, J, S |
| H$_2$ 1–1 $S$(17) | 3.698368 | ⋯ | ⋯ | ⋯ | J | J |
| H$_2$ 2–1 $O$(6) | 3.723689 | C, J | ⋯ | J | C, J, S | C, J |
| H$_2$ 0–0 $S$(14) | 3.724426 | C, J | ⋯ | J | C, J, S | C, J |
| H$_2$ 1–1 $S$(16) | 3.760418 | ⋯ | ⋯ | ⋯ | ⋯ | J |
| H$_2$ 1–0 $O$(7) | 3.807419 | C, J | C, J, S | C, J | C, J, S | C, J, S |
| H$_2$ 0–0 $S$(13) | 3.846113 | C, J, S | C, J, S | C, J | C, J, S | C, J, S |
| H$_2$ 1–1 $S$(14) | 3.941609 | C | ⋯ | J | J | C, J |
| H$_2$ 0–0 $S$(12) | 3.996147 | C, J | C, J | J | C, J, S | C, J, S |
| H$_2$ 2–1 $O$(7) | 4.054046 | C, J | C, J, S | C, J, S | C, J, S | C, J, S |
| H$_2$ 1–1 $S$(13) | 4.067618 | C, J | ⋯ | J | J | C, J |
| H$_2$ 1–0 $O$(8) | 4.162425 | C, J | J | J | J,S | C, J, S |
| **H$_2$ 0–0 $S$(11)** | 4.181077 | C, J, S | C, J, S | C, J | C, J, S | C, J, S |
| H$_2$ 0–0 $S$(10) | 4.409791 | C, J | C, J, S | J | J,S | C, J |
| H$_2$ 1–1 $S$(11) | 4.416611 | J | ⋯ | J | J | J |
| H$_2$ 0–0 $S$(9) | 4.694614 | C, J, S | C, J, S | J | J | J |
| H$_2$ 1–1 $S$(9) | 4.954095 | C, J | C, J, S | J | J | J |
| H$_2$ 0–0 $S$(8) | 5.053115 | C, J, S | C, J, S | C, J, S | C, J, S | J |
| H i Pf$\gamma$ | 3.740576 | ⋯ | J | C | ⋯ | ⋯ |
| **H i $\alpha$** | 4.052269 | C, J | C, J, S | C, J, S | C, J, S | C, J, S |
| [Fe ii] | 3.391896 | ⋯ | ⋯ | ⋯ | ⋯ | ⋯ |
| [Fe ii] | 4.076319 | ⋯ | J | J | J | ⋯ |
| [Fe ii] | 4.081957 | ⋯ | J | J | C, J | C, J |
| **[Fe ii]** | 4.114994 | C, J | C, J | C, J | C, J, S | J |
| [Fe ii] | 4.130183 | ⋯ | ⋯ | ⋯ | J | ⋯ |
| [Fe ii] | 4.434834 | ⋯ | C, J | ⋯ | ⋯ | ⋯ |
| **[Fe ii]** | 4.889139 | J | C, J, S | J | J | ⋯ |

**Note.** For each protostar, the letters C, J, and S indicate that a line is seen in the center, jet, or scattered light aperture, respectively. The lines marked with boldface are the lines selected to make the maps displayed in Figures 2–6.

## Appendix D
## Spectra of Background Stars

In Figures 8–10, we display the native and background-subtracted integrated spectra from apertures centered on stellar continuum sources found in the fields of IRAS 16253, B335, and IRAS 20126. The coordinates of the background stars are listed in Table 7. The background star in the field of HOPS 370 is too saturated to extract a meaningful spectrum. Each aperture is 0″.3 in radius, and the annuli have inner and outer radii of 0″.4 and 0″.5, respectively.

The background-subtracted spectrum for the star in the field of IRAS 16253 shows pure continuum emission plus ice absorption. There is a broad bump at 3.3 μm in the background of IRAS 16253 that is subtracted out, which we attribute to PAH emission (Puget et al. 1985). The star in the field of B335 shows the same except for an emission line at 3.92 μm that is likely an artifact. Both continuum sources in the field of IRAS 20126 show Brα emission, indicating that they are likely to be YSOs.

**Table 7**
Coordinates of the Background Point Sources in the IPA Sample

| Point Source | R.A. (deg) | Decl. (deg) |
|---|---|---|
| IRAS 16253 | 247.090194 | −24.605917 |
| B335 | 294.252906 | 7.569441 |
| IRAS 20126-N | 303.608396 | 41.226213 |
| IRAS 20126-S | 303.608446 | 41.225496 |

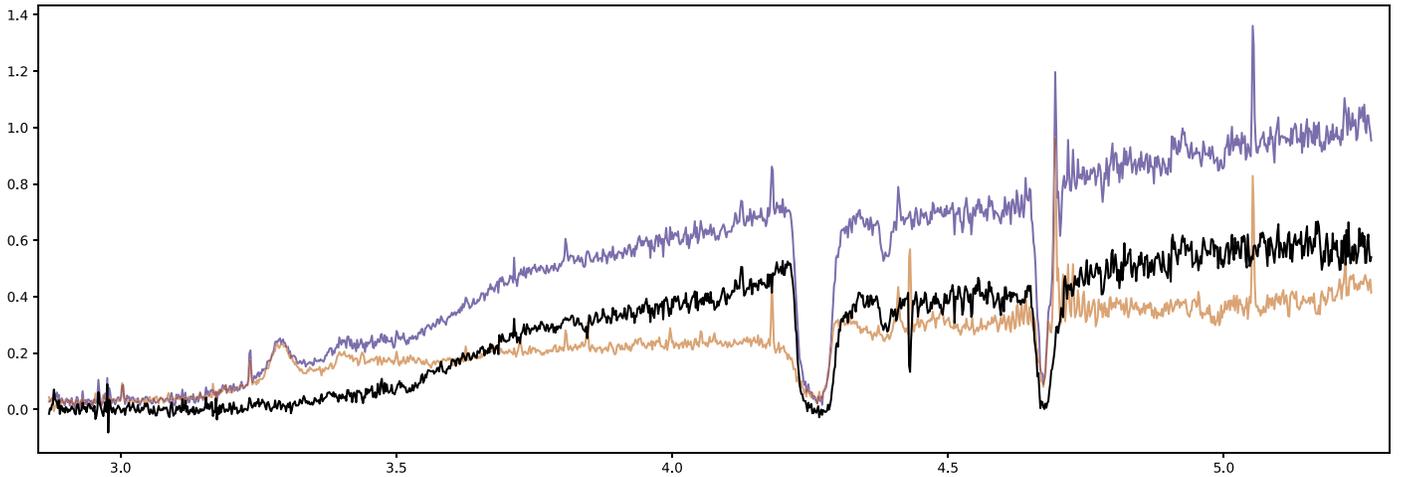

**Figure 8.** Spectrum from the background star in the field of IRAS 16253. The unsubtracted spectrum is shown in blue, the background spectrum is shown in red, and the background-subtracted spectrum is shown in black. The H₂ lines remaining in the subtracted spectrum are likely due to spatial variation in the emission resulting in an imperfect subtraction.

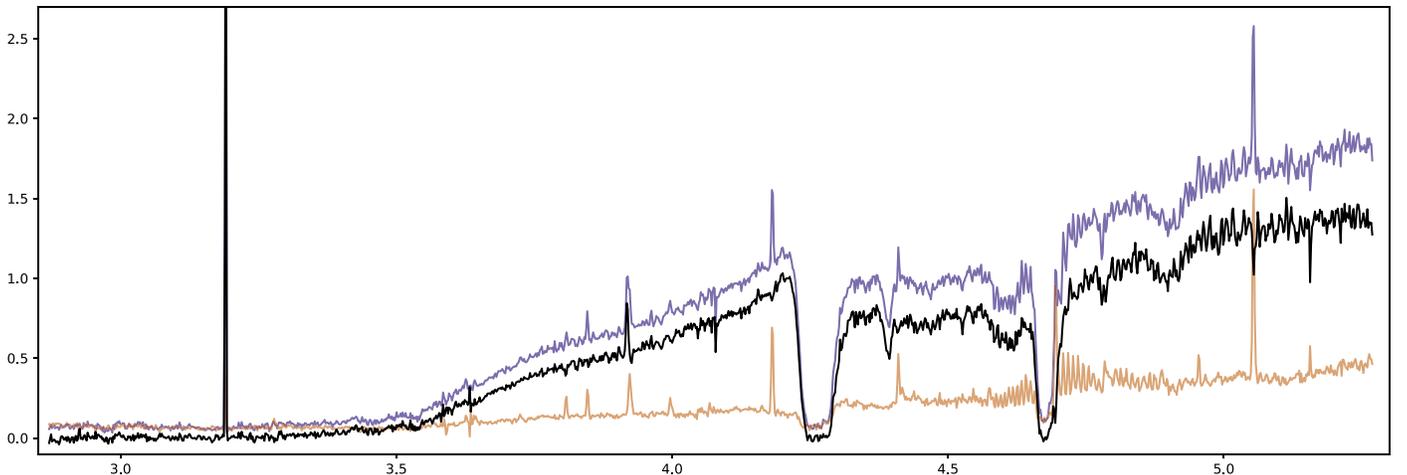

**Figure 9.** Spectrum from the background star in the field of B335. The unsubtracted spectrum is shown in blue, the background spectrum is shown in red, and the background-subtracted spectrum is shown in black. The large spike around 3.2 μm is an artifact, and the H₂ lines remaining in the subtracted spectrum are likely due to spatial variation in the emission resulting in an imperfect subtraction.

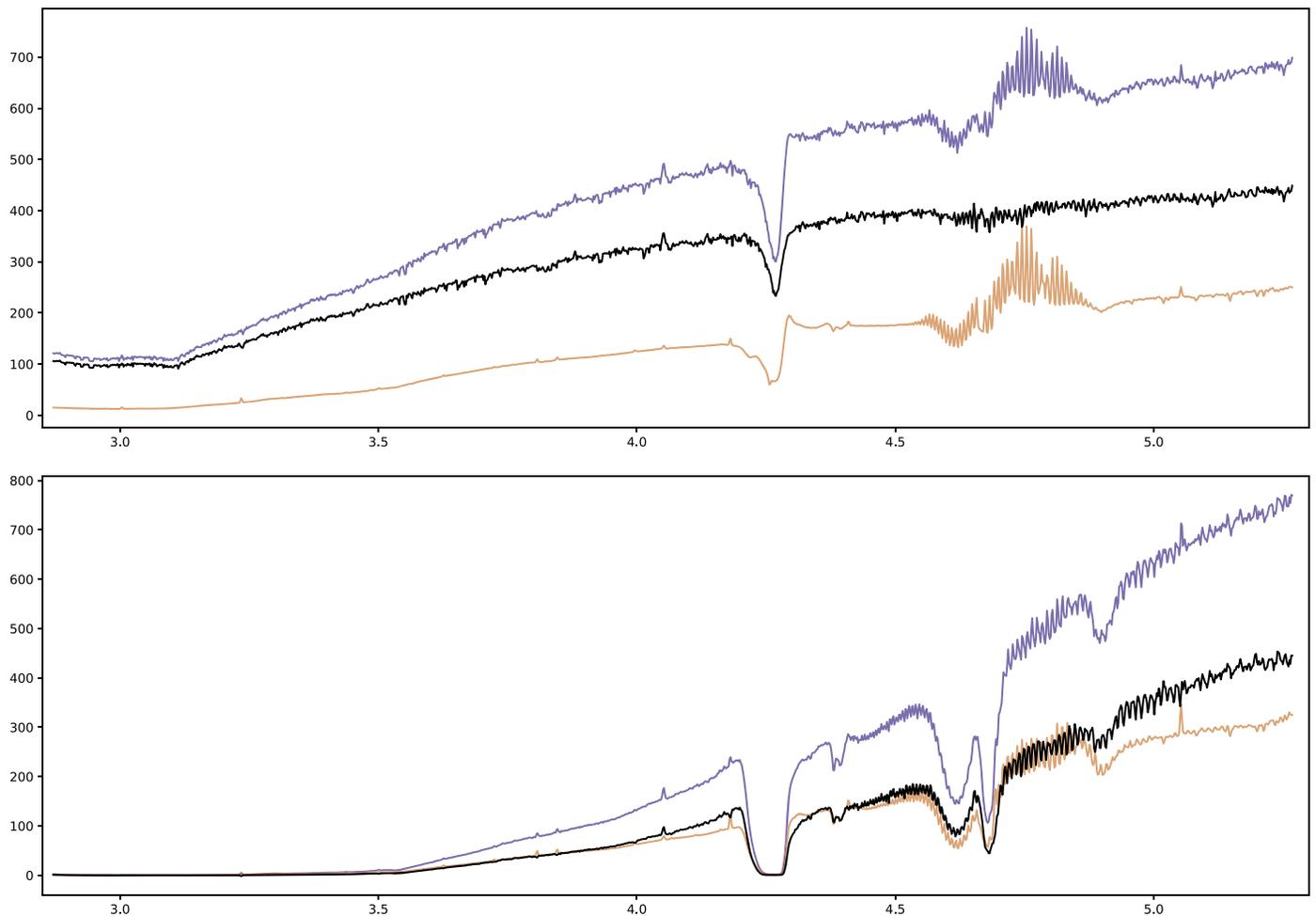

**Figure 10.** Spectra from the background YSOs in the field of IRAS 20126. The top spectrum is from the northern YSO, and the bottom spectrum is from the southern YSO. The unsubtracted spectrum is shown in blue, the background spectrum is shown in red, and the background-subtracted spectrum is shown in black. The unsubtracted spectra both show a clear detection of H I Brα emission at 4.052 μm. The H₂ lines remaining in the subtracted spectrum are likely due to spatial variation in the emission resulting in an imperfect subtraction.

## Appendix E
## Identifying Bona Fide 4.115 μm Iron Emission from IRAS 20126

In Figure 11, we show the continuum-subtracted spectra of the 4.115 μm [Fe II] line from the apertures marking bona fide emission as shown in the top right panel of Figure 6, as well as examples of the large regions of high noise to the east and west that hinder our ability to subtract out the continuum in the cases of a weak or nonexistent signal seen against a complex, bright baseline. The right panels of the figure show that the integrated spectra from the bona fide emission regions marked in the line map show a clear peak around the central wavelength of the line (with some small red- or blueshift), while the integrated spectra from the artifact positions show no peak and show highly structured baselines. We confirm the location of the bona fide [Fe II] emission by comparing to the contours from the 5.34 $\mu$m [Fe II] map from the IPA MIRI data (private communication). There is an apparent line of emission crossing the image from southeast to northwest below the bona fide [Fe II] emission, but we identify this as an artifact as well because it falls outside the MIRI contours.

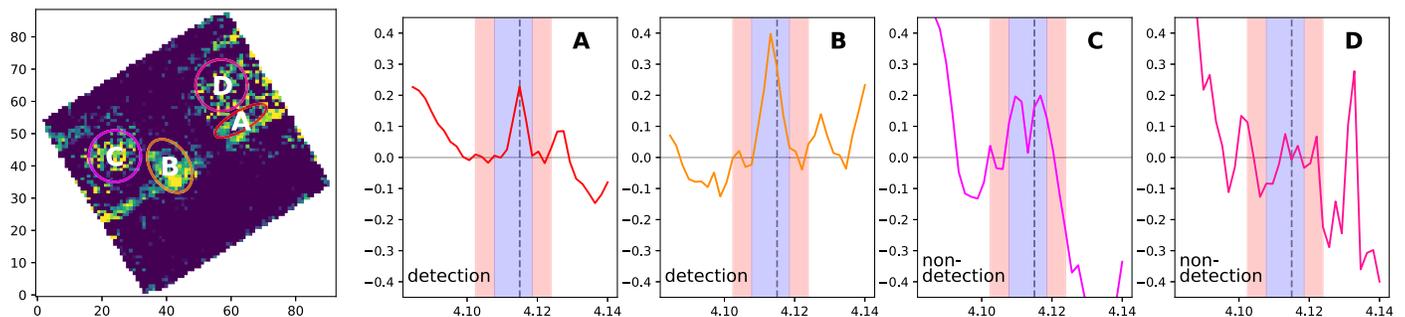

**Figure 11.** Left: continuum-subtracted map of the 4.115 μm [Fe II] line from IRAS 20126, integrated over the line width. The regions of bona fide emission are contained in the orange and red ellipses, while the pink and magenta circles contain artifact emission as shown in the spectra in the right panels. Right: continuum-subtracted spectra centered on the 4.115 μm [Fe II] line from IRAS 20126, integrated over the apertures shown in the left panel. The central wavelength is marked by a black dashed line; the line width integrated to make the image is shown by the blue shaded box, and the red shaded boxes show the wavelength ranges used to estimate the continuum. The gray line at y = 0 demonstrates the baseline determined from a linear fit to the unsubtracted data in the wavelength range of the red shaded boxes. The regions of bona fide [Fe II] emission show a clear peak around the line center, in contrast with the artifact regions.

## Appendix F
## Knot Positions

In Table 8, we list the coordinates of the [Fe II] knot centers, as well as the offsets to the knot centers in the other tracers associated with the jet. We also list the rotation angle used to align the jets vertically for the analysis. An example of the method for determining the distance along the jet axis used to calculate timescales of knots is shown in Figure 12.

**Table 8**
Locations of [Fe II] Jet Knot Centers, Offsets to Jet Knot Centers of Other Tracers, and Rotation to Align Jet Axes Vertically

| Jet Knot[a] | [Fe II] (R.A., Decl.) (deg) | Brα δ$_{R.A.}$, δ$_{decl.}$ (arcsec) | H$_2$ δ$_{R.A.}$, δ$_{decl.}$ (arcsec) | CO δ$_{R.A.}$, δ$_{decl.}$ (arcsec) | Rotation[b] (deg) |
|---|---|---|---|---|---|
| IRAS 16253-N | 247.090493, −24.605997 | −0.150, −0.040 | ⋯ | ⋯ | −20 |
| B335-E E | 294.254323, 7.569294 | −0.079, −0.007 | +0.091, +0.042 | +0.084, +0.011 | −90 |
| B335-E W | 294.254095, 7.569294 | −0.009, −0.007 | ⋯ | ⋯ | ⋯ |
| HOPS 153-NW | 84.487315, −7.115347 | ⋯ | ⋯ | ⋯ | +55 |
| HOPS 153-SE E | 84.487757, −7.115688 | ⋯ | ⋯ | ⋯ | ⋯ |
| HOPS 153-SE W | 84.487647, −7.115612 | ⋯ | ⋯ | ⋯ | ⋯ |
| HOPS 370-N N | 83.865166, −5.159336 | +0.026, +0.009 | +0.063, +0.138 | +0.037, +0.108 | −5 |
| HOPS 370-N S | 83.865161, −5.159438 | +0.007, +0.051 | +0.027, +0.044 | −0.005, +0.009 | ⋯ |
| IRAS 20126-NW W | 303.607827, 41.225980 | ⋯ | +0.052, +0.024 | +0.069, −0.021 | +58 |
| IRAS 20126-NW E | 303.608047, 41.225924 | ⋯ | +0.041, +0.032 | +0.176, +0.190 | ⋯ |

**Notes.**
[a] Individual knots are designated N(orth), S(outh), E(ast), and W(est) when there are two knots in a jet.
[b] Rotation direction is east of north (i.e., counterclockwise).

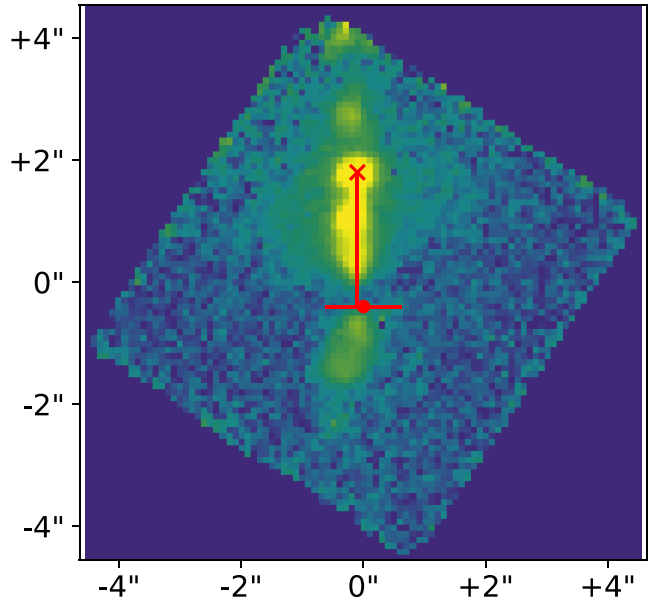

**Figure 12.** An example of the method to determine the distance along the jet axis used to calculate the timescales of knots. Here the B335 [Fe II] 4.889 $\mu$m map is rotated so the jet is vertical. The distance along the jet axis is from the knot center, marked by a red cross, to the point where the vertical line of the jet axis (shown in red) intersects with a horizontal line (in red) drawn from the millimeter position of the protostar, marked by a red circle. *X*- and *Y*-axis values are for the offset from the image center in arcseconds.

## ORCID iDs

Samuel A. Federman  https://orcid.org/0000-0002-6136-5578
S. Thomas Megeath  https://orcid.org/0000-0001-7629-3573
Adam E. Rubinstein  https://orcid.org/0000-0001-8790-9484
Robert Gutermuth  https://orcid.org/0000-0002-6447-899X
Mayank Narang  https://orcid.org/0000-0002-0554-1151
Himanshu Tyagi  https://orcid.org/0000-0002-9497-8856
P. Manoj  https://orcid.org/0000-0002-3530-304X
Guillem Anglada  https://orcid.org/0000-0002-7506-5429
Prabhani Atnagulov  https://orcid.org/0000-0002-4026-126X
Henrik Beuther  https://orcid.org/0000-0002-1700-090X
Tyler L. Bourke  https://orcid.org/0000-0001-7491-0048
Nashanty Brunken  https://orcid.org/0000-0001-7826-7934
Alessio Caratti o Garatti  https://orcid.org/0000-0001-8876-6614
Neal J. Evans, II  https://orcid.org/0000-0001-5175-1777
William J. Fischer  https://orcid.org/0000-0002-3747-2496
Elise Furlan  https://orcid.org/0000-0001-9800-6248
Joel D. Green  https://orcid.org/0000-0003-1665-5709
Nolan Habel  https://orcid.org/0000-0002-2667-1676
Lee Hartmann  https://orcid.org/0000-0003-1430-8519
Nicole Karnath  https://orcid.org/0000-0003-3682-854X
Pamela Klaassen  https://orcid.org/0000-0001-9443-0463
Hendrik Linz  https://orcid.org/0000-0002-8115-8437
Leslie W. Looney  https://orcid.org/0000-0002-4540-6587
Mayra Osorio  https://orcid.org/0000-0002-6737-5267
James Muzerolle Page  https://orcid.org/0000-0002-5943-1222
Pooneh Nazari  https://orcid.org/0000-0002-4448-3871
Riwaj Pokhrel  https://orcid.org/0000-0002-0557-7349
Rohan Rahatgaonkar  https://orcid.org/0000-0002-5350-0282
Will R. M. Rocha  https://orcid.org/0000-0001-6144-4113
Patrick Sheehan  https://orcid.org/0000-0002-9209-8708
Katerina Slavicinska  https://orcid.org/0000-0002-7433-1035
Thomas Stanke  https://orcid.org/0000-0002-5812-9232
Amelia M. Stutz  https://orcid.org/0000-0003-2300-8200
John J. Tobin  https://orcid.org/0000-0002-6195-0152
Lukasz Tychoniec  https://orcid.org/0000-0002-9470-2358
Ewine F. Van Dishoeck  https://orcid.org/0000-0001-7591-1907
Dan M. Watson  https://orcid.org/0000-0001-8302-0530
Scott Wolk  https://orcid.org/0000-0002-0826-9261
Yao-Lun Yang  https://orcid.org/0000-0001-8227-2816